# A Comprehensive multi-species comparison of rotational temperature probes in a DC Ar/$N_2$ micro-hollow cathode discharge

**Dimitrios Stefas, Belkacem Menacer, Alice Remigy, Nikolaos Chazapis, Guillaume Lombardi, Claudia Lazzaroni, Kristaq Gazeli***

Laboratoire des Sciences des Procédés et des Matériaux (LSPM – CNRS), Université Sorbonne Paris Nord, Villetaneuse, UPR 3407, F-93430, France

*emails: kristaq.gazeli@lspm.cnrs.fr

**ORCID**

0000-0003-2390-4603 (D.S)
0009-0006-1224-6683 (B.M)
0000-0003-2743-2981 (A.R)
0009-0001-2307-1106 (N.C.)
0000-0001-6583-0046 (G.L)
0000-0003-0966-560X (C.L)
0000-0002-5479-0373 (K.G.)

## ABSTRACT

Accurate gas temperature ($T_{Gas}$) determination in microplasmas is critical for optimizing their applications, yet isolated diagnostic approaches may yield misleading results, especially under strong non-equilibrium conditions. Here, high resolution rotational spectra of $N_2$(C), OH(A), NH(A) and NO(A), generated in the plasma jet of a DC Ar/$N_2$ microhollow cathode discharge (MHCD), are recorded and their associated rotational temperatures ($T_{rot}$) are cross compared. A detailed experimental analysis and robust fitting of the rotational spectra are performed, achieving a reliable estimation of $T_{Gas}$. The dominant formation mechanisms of these species and their corresponding impact on rotational population distributions are also interrogated. Particularly, our findings indicate that the $T_{rot}$ of $N_2$(C) is significantly influenced by energy transfers from argon metastables ($Ar^m$) and spectral interference from NH(A). This makes it unreliable as a thermometric probe in Ar-rich MHCD, unless complex analyses are employed. In contrast, OH(A) rotational population distribution appears to be less sensitive to Ar-induced perturbations across various discharge currents and pressures, providing more straightforward results. For all molecules considered, this study reveals the conditions under which all the measured $T_{rot}$ can be reliably considered to be in equilibrium with $T_{Gas}$. This highlights the importance of cross-validating multiple thermometric probes and investigating relevant excitation kinetics when measuring $T_{rot}$ in reactive microplasmas.

## INTRODUCTION

Micro Hollow Cathode Discharges (MHCDs) have emerged as a promising research subject in the plasma physics and applications community since their initial investigation in the 1990s [1]. They represent a class of microplasmas since at least one spatial dimension lies in the micrometer scale. Several MHCD arrangements have been proposed to generate discharges in various gases, driven by different types of high voltage waveforms [1–8]. Among them, a distinct design involves a tri-layer structure consisting of two electrodes separated by a dielectric material, usually driven by a DC high voltage [1, 2, 9]. A central hole with a diameter of a few hundreds of micrometers, penetrates all three layers. This design is also characterized by a small thickness of the dielectric material (<1 mm). Another interesting aspect of this assembly is that it can be placed in the junction of two isolated chambers which communicate through the hole and can have either the same or different pressures [6, 7, 10]. In the former case, the plasma remains confined in the hole and covers a portion of the cathodic surface, while in the latter case a luminous plasma jet is formed in the low-pressure chamber due to the pressure difference between the two chambers. These MHCDs consume very low electric power (typically few W) and operate from moderate (few mbar) to high (100s mbar) pressures. Additionally, due to the small diameter of the hole, the deposited power density reaches high values (kW/cm$^3$) which facilitates the generation of high electron densities (up to $10^{15}$ $cm^{-3}$) at relatively low operating DC voltages (∼1 kV) [11].

Furthermore, MHCDs remain out of thermodynamic equilibrium by presenting electron temperature ($T_e$) significantly higher than the neutral gas temperature ($T_{gas}$). This enables efficient energy transfer to atoms and molecules inducing various ionization, excitation and dissociation processes, and thus enhancing the chemical reactivity without excessive heating of the bulk gas. This is particularly interesting for molecular nitrogen dissociation, which requires a significant energy to be split (9.8 eV) [12–14]. Thanks to their non-equilibrium nature, MHCDs can be used as sources of reactive N-atoms at relatively low gas temperatures ($T_{gas}$<1000 K), which makes them energetically efficient for the synthesis of various nitride compounds such as amorphous or hexagonal boron nitride (a-BN and h-BN, respectively) [15, 16]. This is an advantage compared to traditional high-temperature synthesis methods of BN. A notable example is the synthesis of BN thin films on 2–inch silicon substrates using a ns-pulsed MHCD for an energy cost lower than traditional high temperature synthesis methods [15]. The relatively small gas temperature of MHCD also favors their interaction with heat sensitive substrates. In film synthesis applications, $T_{gas}$ affects the efficiency of the precursor decomposition efficiency into reactive species and their mobility over the surface of different substrates. Besides, as $T_{gas}$ refers to the translational energy of the neutrals in the plasma, it controls the frequency of collisions between different species and, thus, the likelihood of appearance of key chemical reactions. Consequently, knowledge of $T_{gas}$ is indispensable for the development and refinement of MHCD-based processes as well as the validation of relevant kinetic models [17, 18].

However, accurate determination of $T_{gas}$ in MHCDs is not an easy task. Optical emission spectroscopy (OES) is a reliable non-invasive diagnostic to estimate $T_{gas}$ in many practical cases. For instance, common molecular emissions adopted for this purpose include NO(A–X), OH(A–X), $N_2$(C–B), $N_2^+$(B–X), $N_2$(B–A), NH(A–X), as well as other diatomic molecules (and/or atoms) that may be present depending on the gas mixture and discharge conditions [19–23]. Particularly, the rotational temperature ($T_{rot}$) of excited molecular states is obtained by fitting the experimentally-recorded high-resolution molecular bands with corresponding synthetic spectra [19, 24–28]. This method requires identical rotational distributions between the ground and excited states and, thus, the $T_{rot}$ of the excited state is in thermal equilibrium with the translational temperature ($T_{gas}$) of the neutrals. Consequently, the rotational distribution can be simply described by a single rotational temperature. This is true especially in non-thermal plasmas where $T_e$ is significantly higher than $T_{gas}$ and the collisional frequency is high enough to allow rotational population equilibrium establishment between the ground and the excited state. However, depending on the molecule and its excitation mechanisms, an isolated analysis based on a single $T_{rot}$ could not be sufficient to describe the rotational distribution population of the excited state. In this case, fitting the spectra with a single $T_{rot}$ would only provide an average temperature between a higher and a lower value which would not correspond to the actual $T_{gas}$. Consequently, this $T_{rot}$ can only provide an upper limit of the actual $T_{gas}$ and more sophisticated analyses are needed to achieve an accurate assessment [19].

The present work focuses on an argon/nitrogen (Ar/$N_2$) MHCD driven by a DC high voltage. This microplasma source represents an ideal reactive environment to study the $T_{rot}$ variations of the excited states of different molecular bands (NO(A), OH(A), $N_2$(C)) formed. Their detailed analysis allows for investigating different formation mechanisms of the upper states and making better estimations of $T_{gas}$. This is achieved through a robust cross-comparison of the different $T_{rot}$ extracted under representative experimental conditions. Section II refers to the description of the microplasma source, experimental setup, discharge diagnostics and data analysis methods applied. Section III contains the main results obtained, followed by corresponding discussions in section IV. The main conclusions derived are given in section V.

# EXPERIMENTAL SETUP

## MHCD description

The MHCD studied in the present work is an anode-dielectric-cathode stack consisting of two 100 µm thick layers of molybdenum (Mo) electrodes separated by a 750 µm thick layer of Alumina ($Al_2O_3$). These three layers are glued together with an epoxy glue. A 400 µm diameter hole is laser-drilled at the center of the stack and traverses all the three layers. The MHCD stack is illustrated in **Fig.1a**.

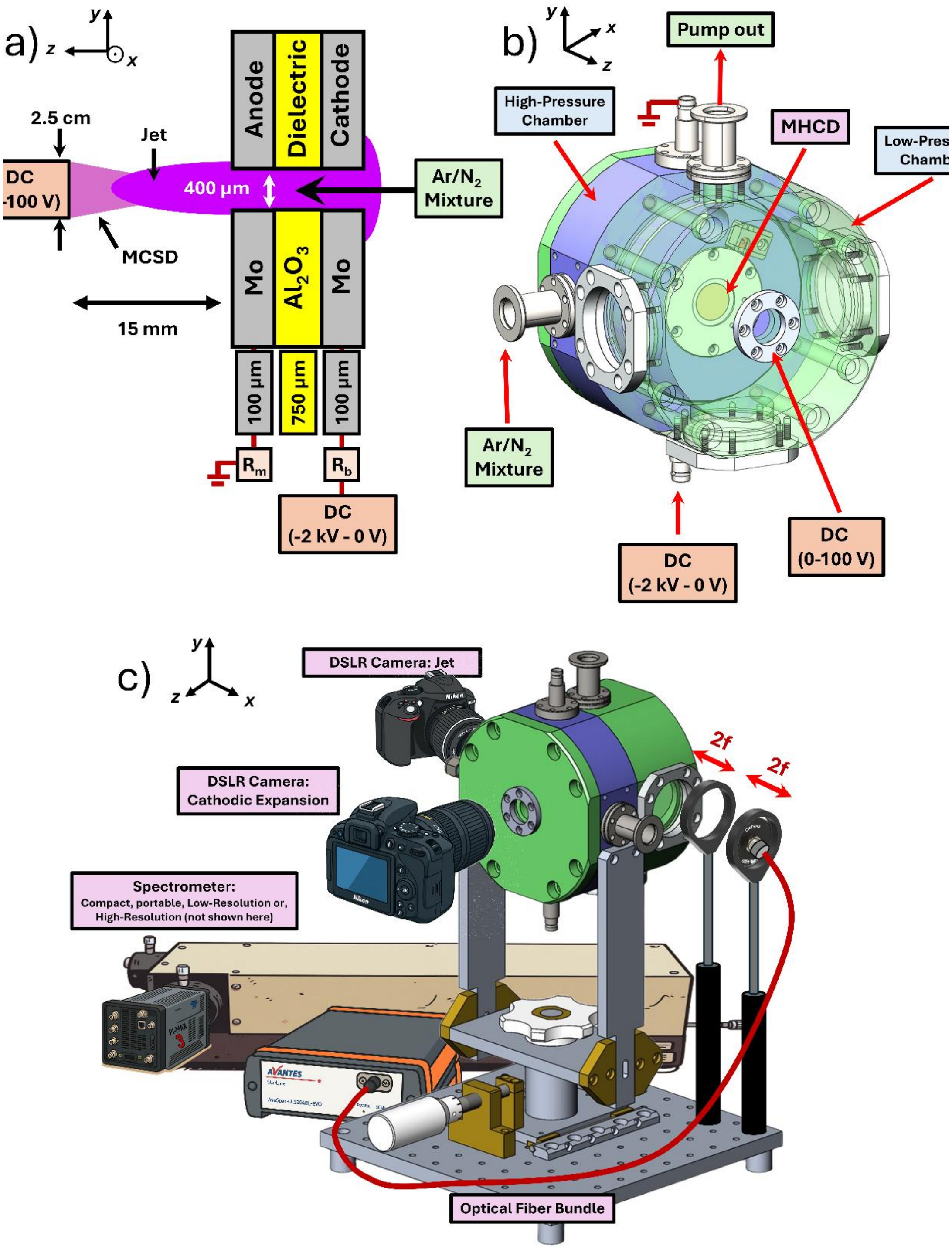

**Fig.1**. **a**) Schematic of the MHCD assembly facing a substrate holder (not drawn in scale). **b**) MHCD reactor (the substrate holder is not shown in the figure for clarity reasons). **c**) Experimental setup used for OES and broadband conventional imaging measurements.

The MHCD assembly (**Fig.1a**) is positioned inside a sealed reactor composed of a low– and high–pressure chamber (see **Fig.1b**). An Ar/$N_2$ gas mixture (Air Liquide; Alphagaz 1 for Ar and Alphagaz 2 for $N_2$) is introduced into the high–pressure chamber, both flow rates being regulated using dedicated mass flow controllers (GF-40, Brooks Instrument). The pressure difference between the two chambers induces a gas jet through the hole which expands in the low-pressure chamber. This chamber contains two windows that allow for optical accesses in the plasma jet formed when the discharge is turned "ON" (see **Fig.1b** and **1c**). The specifications of the gas cylinders used are given in **Table 1**. The two chambers are separated by the MHCD assembly, and their operating pressures are set as follows: $P_{high}$=20–80 mbar (high pressure side) and $P_{low}$=10 mbar (low pressure side). The gas pressures in both chambers are measured using dedicated pressure gauges (low pressure chamber: CCR 363, Pfeiffer; high pressure chamber: CCR 361, Pfeiffer). These are regulated by adjusting the gas fluxes via the mass flow controllers and a manual valve before a vacuum pump (UNO 6, Pfeiffer). The electrodes of the MHCD are driven by a negative DC power supply that can deliver high voltages down to -2 kV (T1070006, THQ, ISEG). The electrode found on the low-pressure side is connected to the ground representing the anode, the other one in the high-pressure side represents the cathode. The DC negative high voltage is supplied through a ballast resistor ($R_b$ = 480 kΩ) limiting the current, avoiding arcing and thereby damage to the source. Just after $R_b$, a voltage probe (Teledyne LeCroy HVP120; attenuation ratio 100:1) is connected for measuring the discharge ignition voltage ($V_d$). The discharge current $I_d$ is obtained by connecting another voltage probe (Rohde & Schwarz RT-ZP10; attenuation ratio 10:1) just after another resistor $R_m$ = 180 Ω and measuring the voltage drop on it ($V_i$). A third electrode (2.5 cm diameter) is installed in the low-pressure chamber 15 mm away from the MHCD anode. This electrode is positively polarized and can reach up to +100 V using another DC power supply (T1070006, THQ, ISEG). This helps in expanding the plasma jet towards the biased electrode. This configuration is called micro cathode sustained discharge (MCSD) and mimics the operation of a larger MHCD reactor used for nitride deposition applications [15, 16].

**Table 1**. Specifications of the gas cylinders used

| Cylinder | Gas | Purity (%) | $H_2O$ (ppm) | $C_nH_m$ (ppm) | $O_2$ (ppm) | $CO_2$ (ppm) | CO (ppm) | $H_2$ (ppm) |
|---|---|---|---|---|---|---|---|---|
| Alphagaz 1 | Ar | 99.999 | 3 | 0.5 | 2 | - | - | - |
| Alphagaz 2 | $N_2$ | 99.9999 | 0.5 | 0.1 | 0.1 | 0.1 | 0.1 | 0.1 |

## Broadband Conventional Imaging and Optical Emission Spectroscopy (OES)

Conventional photos of the visual appearance of the plasma are recorded using a Digital Single Lens Reflex (DSLR) camera (NIKON D5300 coupled with NIKKOR 18-105 mm f/3.5-5.6G ED VR lens), having 1/15 s integration time and ISO 100 (see **Fig.1c**). Furthermore, the wavelength-integrated plasma emission is collected using a 100 mm plano convex lens (LA4380, Thorlabs) and fed into a quartz optical fiber bundle (FCUV400–2, Avantes) using a 2f–2f focusing configuration between the plasma and the fiber. With this configuration an inverted image, the same size as the plasma, is focused on the plane where the fiber is positioned, offering the possibility for spatially resolved measurements. This is done by moving the position of the fiber along the imaged axis of the plasma using a micrometric translation stage. The other end of the fiber is coupled to the entrance slit of a compact portable spectrometer (AvaSpec–ULS4096CL-EVO, Avantes). It has a 75 mm focal length and a fixed slit width of 10 μm. It is equipped with a 300 lines/mm diffraction grating, and a 4096-pixel CMOS linear image sensor, leading to a resolution of ~0.5 nm/pixel for a spectral range between 200 and 1100 nm. Thus, it allows recording broad emission spectra of the plasma jet with low spectral resolution. Additionally, a high-resolution spectrometer (THR 1000, Jobin Yvon; 1 m focal length) is used to record specific molecular bands and infer the $T_{rot}$ of their excited states. It is equipped with a 1200 lines/mm grating blazed at 500 nm. To record the plasma-emitted radiation, an ICCD camera (PI-MAX 3 1024i, Princeton Instruments) is used, with a 12.8×12.8 μm$^2$ pixel size, leading to an overall monochromator resolution of ~0.005 nm/pixel (see **Fig.1c**). The spectral response of the two spectrometers is corrected using a Halogen-Deuterium calibration lamp (DH3-PLUS-CAL, OceanOptics). The corrected spectrum is obtained as:

$$I_{corr}(\lambda) = I_{raw}(\lambda)/R(\lambda) \quad \mathbf{(1)}$$

, where $R(\lambda)$ is the response function of the instrument. The setup is schematically shown in **Fig.1c**.

To determine the $T_{rot}$ of different excited molecules formed, the recorded rotational emission spectra are fitted with simulated/synthetic spectra generated by a custom-made code [27–29], as well as using the software MassiveOES [24–26] and the Moose Python library [30]. MassiveOES offers line lists for $N_2$(C-B), OH(A-X) and NH(A-X), but not for NO(A-X), for which we used the line list offered by ExoMol [31].

We briefly explain here the principle of the method for fitting synthetic spectra.

For a diatomic band, the emission intensity of a single rotational line $J'\rightarrow J''$ (within a fixed vibronic band $v'\rightarrow v''$ and for an optically thin plasma) is:

$$I_{J'J''}(\lambda) \sim N_{J'} A_{J'J''} h\nu \sim g_{J'} S_{J'J''} \frac{e^{-E_{J'}/(k_B T_{rot})}}{Q_{rot}(T_{rot})} \quad \mathbf{(2)}$$

, where $N_{J'}$ is the upper-state rotational population, $A_{J'J''}$ the Einstein coefficient, $\nu$ the line frequency, $g_{J'}=2J'+1$ the rotational degeneracy (modified by spin–orbit/nuclear-spin factors when applicable), $S_{J'J''}$ the Hönl–London factor for the corresponding branch (P, Q, or R), $E_{J'}$ the upper-state rotational energy, and $Q_{rot}$ the rotational partition function. The upper-state rotational energy is:

$$E_{J'} \simeq hc(B'J'(J'+1) - D'(J'(J'+1))^2) \quad (3)$$

, with $B'$ and $D'$ the effective rotational and centrifugal distortion constants of the emitting upper electronic–vibrational state. After spectral response correction, a line-by-line Boltzmann plot can be constructed as:

$$ln(\frac{I_{J'J''}}{R(\lambda)S_{J'J''}g_{J'}v^4}) = -\frac{E_{J'}}{k_B T_{rot}} \quad (4)$$

, so that the slope yields $-1/(k_B T_{rot})$. In MassiveOES/Moose, the entire rotational band is fitted by computing synthetic rotational spectra with the correct $S_{J'J''}$ and $E_{J''}$, scaling them by a Boltzmann distribution at a trial $T_{rot}$, and then convolving with the instrument function to mirror the experimental resolution:

$$I_{syn}(\lambda; T_{rot}) = \left[\sum_{J',J''} I_{J'J''}\delta(\lambda - \lambda_{J'J''})\right] * G_{inst}(\lambda; FWHM_{inst}) \quad (5)$$

The experimental spectrum is then compared to the synthetic one, using a least- squares algorithm that systematically adjusts $T_{rot}$ in the synthetic model to minimize the residual error between simulated and measured spectra, yielding the best- fitted rotational temperature:

$$min \left\| \sum_{\lambda} (I_{corr}(\lambda) - I_{syn}(\lambda; T_{rot}))^2 \right\| \quad (6)$$

For the bands used in the present work, selection rules yield P (ΔJ=−1), Q (ΔJ=0), and R (ΔJ=+1) branches, etc. (see, e.g., **Fig.4**), with state-specific Hönl–London factors $S_{J'J''}$. The correct $S_{J'J''}$ tables for the studied emissions are embedded in the line-strength files used by MassiveOES/Moose and by the custom-made code, ensuring internally consistent weighting across branches. MassiveOES also provides a state-by-state fitting tool, which is convenient to calculate the quantum rotational level populations directly from the rotational spectrum and is also relatively robust to overlapping lines. This provides the so-called Boltzmann plot relatively easily, whereas the traditional manual way of constructing Boltzmann plots from heights of rotational lines is a lengthy task and is limited to non-overlapping lines.

Finally, it is noted that the fidelity of $T_{rot}$ as an estimate for $T_{gas}$ depends on the excitation pathway of the upper state of a probe molecule. Accordingly, in the subsequent sections we compare $T_{rot}$ extracted from OH(A), $N_2$(C), NH(A), and NO(A) bands to achieve more reliable estimations of $T_{gas}$.

# RESULTS

## Discharge morphology: Plasma Jet and Cathodic Expansion

**Fig.2a** shows representative DSLR photographs of the plasma jet emerging from the MHCD hole in the low–pressure chamber. These are shown for different Ar/$N_2$ mixtures and a fixed discharge current of 2.5 mA (first three rows of the figure). Furthermore, similar photos under different currents for a 50% Ar–50% $N_2$ mixture are shown (rows 4 to 7). It should be mentioned that these photos offer qualitative information about the discharge structure since they do not contain emissions below 350 nm due to the limited response of the camera sensor in the UV range. In pure argon (i.e., 100% Ar–0% $N_2$) the jet is very short, faintly violet and confined near the hole exit. Adding 10% $N_2$ in Ar (i.e., 90% Ar–10% $N_2$) highly increases the visible plume length and introduces a strong purple–red tail most probably due to the formation of excited nitrogen bands such as $N_2$ Second Positive System (SPS) in addition to Ar I lines. This length seems to remain constant until a 50%-50% Ar/$N_2$ mixture for which the jet color slightly changes by becoming orange-violet. Maintaining the gas composition at 50% Ar–50% $N_2$ and decreasing the discharge current from 2.5 mA down to 0.5 mA (see bottom five rows in **Fig.2a**) progressively reduces both the length and luminous intensity of the jet. This trend highlights that both nitrogen admixture and discharge current are key parameters for sustaining an MHCD plasma jet. Controlling its length and intensity is important for material synthesis applications where the effluent needs to reach remote substrates while achieving efficient dissociation of molecular precursors [15, 32].

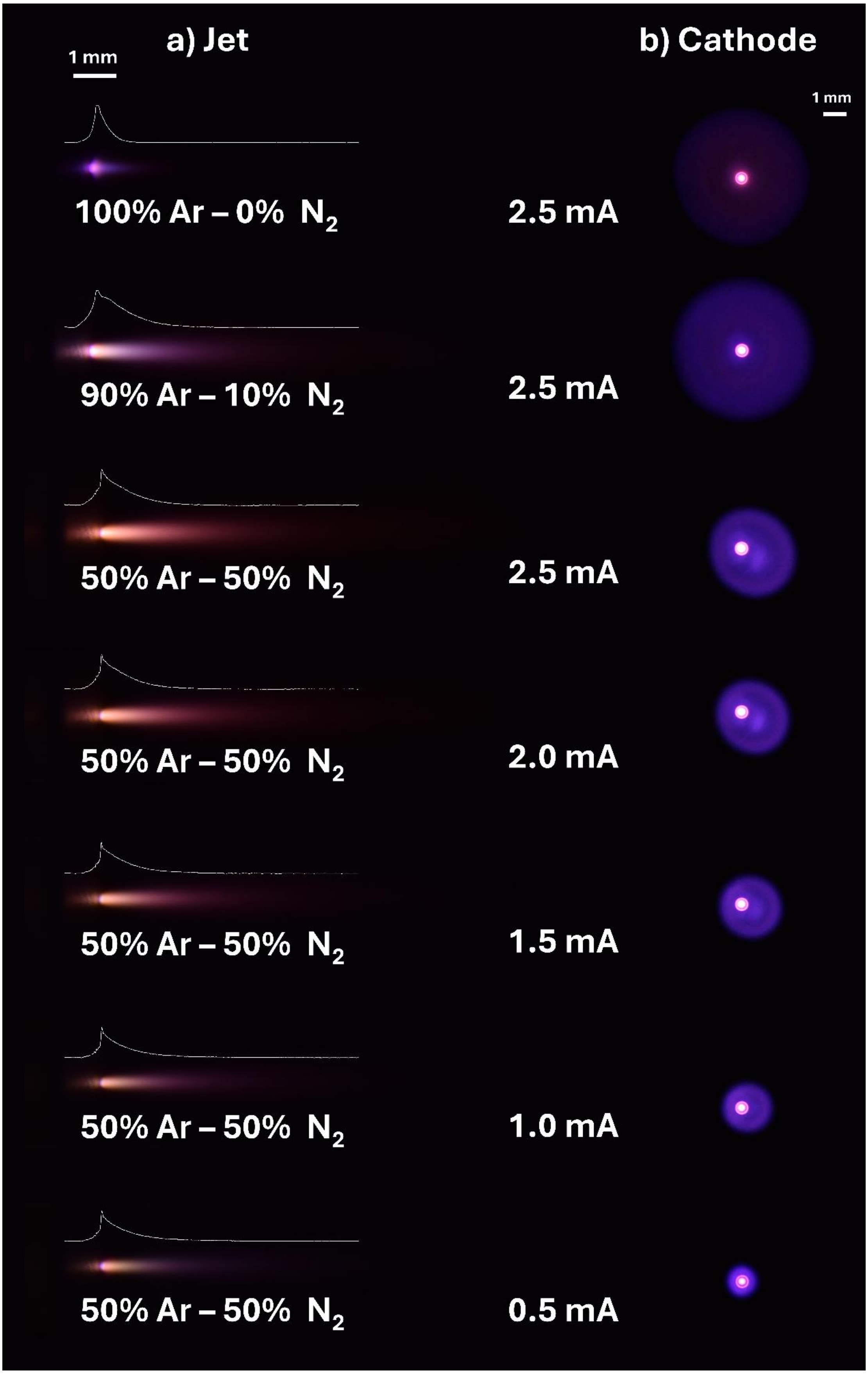

**Fig.2.** Representative DSLR images (integration time: 1/15s, ISO: 100) of the discharge for different operating conditions showing the effect of gas mixture and current on **a**) plasma jet and **b**) cathodic expansion. $P_{high}$=30 mbar and $P_{low}$=10 mbar.

**Fig.2b** shows the morphology of the cathodic expansion for the same experimental conditions as **Fig.2a**. For 100% Ar and $I_d$=2.5 mA, the cathodic glow appears as a small, intense central spot (i.e., the MHCD hole) surrounded by a faint uniform and symmetric violet halo extending up to about ~2 mm away from the hole's center. Introducing only 10% $N_2$ in the gas mixture produces a brighter discharge in the hole surrounded by a diffuse circular halo indicating contributions from molecular nitrogen emissions, as for the jet. In the 50%/50% Ar/$N_2$ mixture at $I_d$=2.5 mA the expansion around the hole becomes smaller, slightly non-symmetric and obtains an elliptical form with clear ring patterns disturbing the intensity uniformity. This non-uniformity and elliptical shape of the emission pattern are also maintained for smaller currents down to 0.5 mA. The decrease of the current also causes a contraction of the cathodic expansion area. The systematic decrease in both size and intensity of the cathodic expansion with lowering current mirrors the behavior of the jet. The addition of large nitrogen concentrations (>10% $N_2$) in the Ar/$N_2$ mixture breaks the symmetric glow profile observed in the first two rows of **Fig.2a**. The origin of this distortion is not entirely clear. However, we believe that the distorted shapes in $N_2$-rich Ar/$N_2$ mixtures are likely the consequence of a discharge constriction coupled to cathode surface inhomogeneities. In the case of pure Ar and its mixtures with small amounts of $N_2$ (up to 10%), electrons will most likely collide elastically with Ar atoms. Thus, they do not lose much energy between collisions with Ar atoms and $N_2$ molecules (which are essentially a minority) until they gain enough energy to ionize the Ar atoms. This leads to an easier expansion on the cathodic surface. However, at high $N_2$ percentages, a critical amount of $N_2$ molecules becomes available for collisions. These will most likely react with electrons which will lose significant amounts of their energy being converted to vibrationally excited states of $N_2$. Thus, the ionization efficiency of the MHCD will be decreased. In order for the discharge to be sustained, it will shrink itself into a smaller region to maintain a high enough current density and temperature for ionizing the gas. Furthermore, Ar ions are heavy and inert. When they bombard the cathode, they physically sputter its surface, removing impurities in a rather uniform way, which maintains a uniform surface for electron emission. On the contrary, nitrogen ions and radicals are reactive and will interact with the cathode in a rather non-uniform way, generating metal nitrides or adsorbed nitrogen layers. This creates cathodic zones with dissimilar secondary electron emission coefficients. This makes the discharge flow in the regions where electrons are easier to be extracted, thus inducing an asymmetric expansion. This also is the cause of a faster degradation of the MHCD when operating in $N_2$-rich mixtures compared to pure Ar, as reported in previous relevant works [6, 10, 17, 22, 23].

## Radiative species detection and variation under different MHCD operating conditions

**Fig.3** shows the emissive species detected in the UV-to-NIR optical emission spectrum of the MHCD jet. **Fig.3a** and **Fig.3b** show the emissions in the spectral ranges 230–480 nm and 600–930 nm, respectively. These spectra are obtained using a gas mixture of 50% Ar and 50% $N_2$, $P_{high}$ = 30 mbar, $P_{low}$ = 10 mbar and $I_d$=2.5 mA. Since the discharge is generated in an Ar/$N_2$ gas mixture, the prominent emissions mainly originate from different nitrogen-based molecular bands, namely, NO($A^2\Sigma^+$–$X^2\Pi$), $N_2$($C^3\Pi_u$–$B^3\Pi_g$) ($N_2$(SPS)) and $N_2$($B^3\Pi_u$–$A^3\Sigma^+_u$) ($N_2$(FPS)), as well as a plethora of Ar I lines. Besides, after careful inspection of the emission spectrum, the emissions of OH($A^2\Sigma^+$–$X^2\Pi$) and $N_2^+$($B^2\Sigma^+$–$X^2\Sigma$) ($N_2^+$(FNS)) are also observed, the former overlapping with $N_2$(SPS) and the latter having a low relative intensity. The formation of NO and OH molecules in the spectrum is a strong indication of $N_2$ dissociation in the discharge and may be due to the presence of water impurities in the MHCD reactor which originate either in the gas bottles (see **Table 1**) and/or the humidity accumulated in the chamber walls. The population of various vibrational levels (v'>0, v''>0) is distinguishable both for the upper and lower electronic states of $N_2$ (up to #6 for the A, #9 for the B, and #3 for the C states) and NO (up to #2 for the A and up to #7 for the X states). It is noted that, for all operating conditions studied, no radiative transition is observed between 480 and 600 nm.

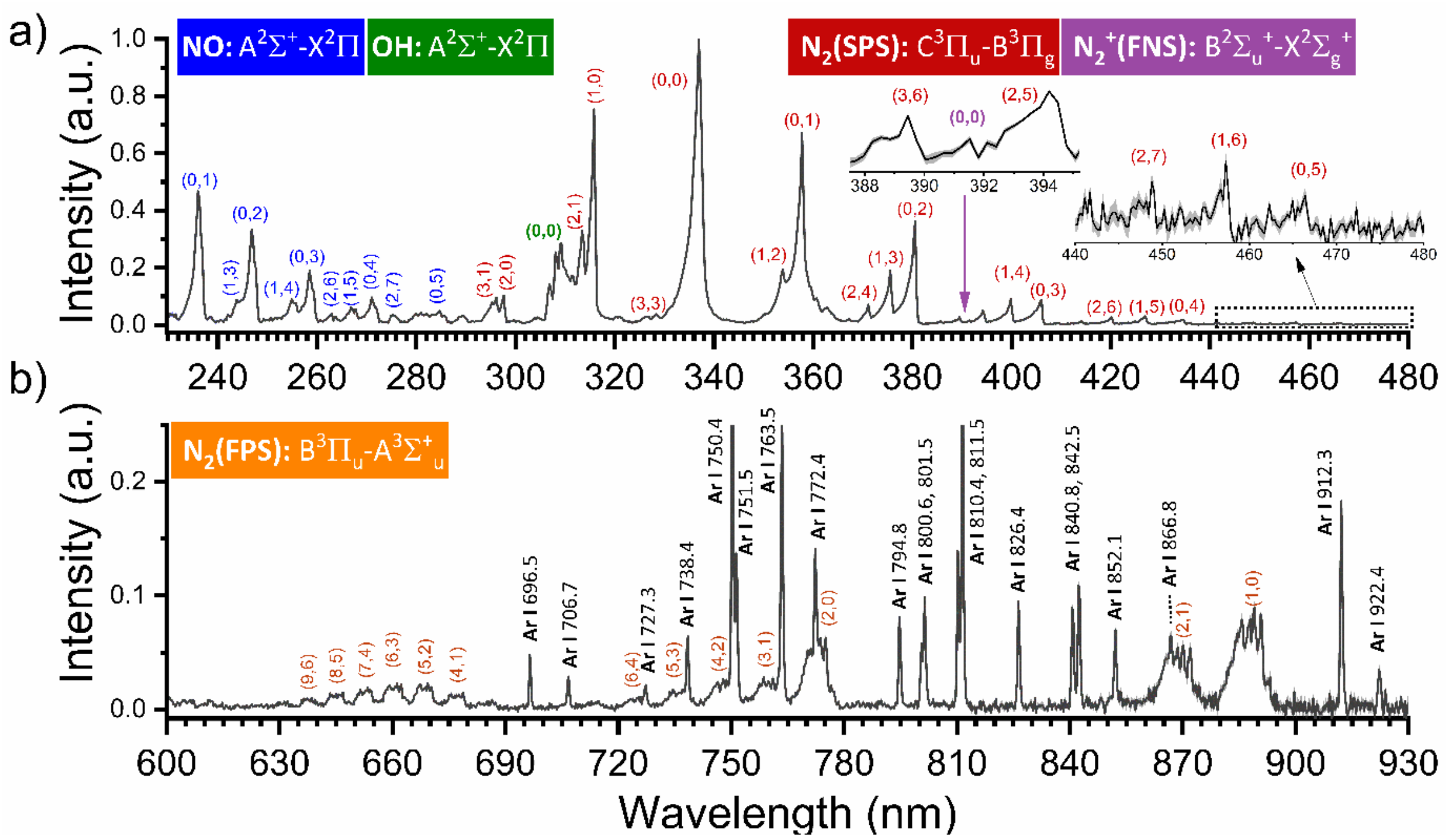

**Fig.3.** Normalized emission spectrum of the MHCD jet (z=0 mm, 50% Ar – 50% $N_2$, $P_{high}$=30 mbar, $P_{low}$=10 mbar, $I_d$=2.5 mA) in the spectral ranges: **a**) 230–480 nm, **b**) 600–930 nm.

**Fig.4** presents high-resolution rotational spectra of selected molecular bands from the broadband spectrum of **Fig.3**. These are accompanied by a detailed identification of the different rotational line transitions detected in each band and are essential for the determination of their corresponding rotational temperatures. **Fig.4a** shows the emission of the OH($A^2\Sigma^+$–$X^2\Pi$) (v'=0, v''=0) band head centered around 309 nm. Similarly, **Fig.4b** shows the $N_2$($C^3\Pi_u$–$B^3\Pi_g$) (v'=0, v''=0) band head centered around 337 nm and **Fig.4c** shows the NO($A^2\Sigma^+$–$X^2\Pi$) (v'=0, v''=1) band head in the 233–237 nm range. In contrast to the broadband spectrum, the high spectral resolution utilized here allows for the clear distinction of the individual rotational lines within the vibrational bands. Selected distinct rotational branches are identified and labelled (e.g., P, Q, R branches for OH; P, R branches for $N_2$ ; P, O branches for NO). Furthermore, the recording of these fine structures with high resolution is essential for the application of optical emission spectroscopy (OES) diagnostics, specifically for the determination of the $T_{rot}$ through the fitting of synthetic spectra to the experimental data.

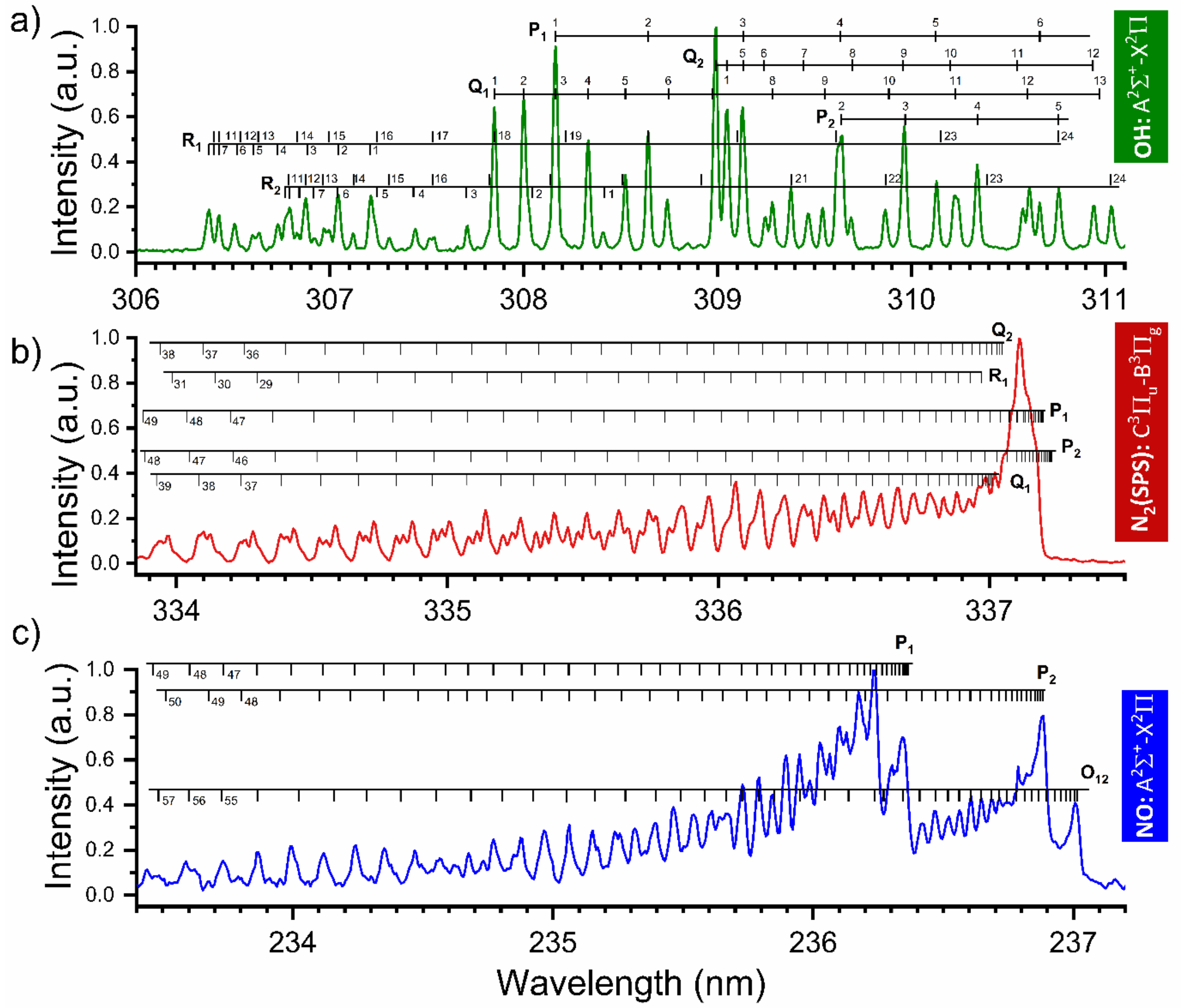

**Fig.4.** High-resolution emission spectra of the MHCD jet ($z$=0 mm, $P_{high}$=30 mbar, $P_{low}$=10 mbar, $I_d$=2.5 mA) showing the characteristic rotational structures of: **a**) OH($A^2\Sigma^+$–$X^2\Pi$) at 100% Ar, **b**) $N_2$($C^3\Pi_u$–$B^3\Pi_g$) at 100% Ar, and **c**) NO($A^2\Sigma^+$–$X^2\Pi$) at 90% Ar – 10% $N_2$.

**Fig.5** shows the emission intensities (using the integrated area under each radiative transition) of various excited species versus the axial distance along the jet for different voltage biases on the third electrode (and otherwise same conditions as **Fig.3**). Each plot (**Fig.5a–f**) corresponds to a different radiative species: **a**) NO(A–X) 230-240 nm, **b**) OH(A-X) 306-311 nm, **c**) $N_2$(C–B, Δv=0) 329-343 nm, **d**) $N_2$(B–A, Δv=3) 640-680 nm, **e**) Ar I ($2p_7$–$1s_5$ and $2p_2$–$1s_3$) near 772.4 nm (these cannot be distinguished with the compact spectrometer), both ending to metastables states, and **f**) Ar I ($2p_2$-$1s_5$) near 696.5 nm, also populating a metastable state. Different colors shown represent different values of the third electrode bias (0 V, white, up to 100 V, red).

At 50% Ar–50% $N_2$ and $I_d$=2.5 mA, all species intensities in **Fig.5** exhibit axial profiles that peak around z=0 mm and then decay by an order of magnitude over the next 8–10 mm, approaching the background level beyond ~12 mm. The third-electrode bias primarily amplifies these profiles without shifting the peak position or changing the profile shape. Specifically, when moving from 0 to 100 V, the peak intensity increases by ~20–30% for NO(A–X) (**Fig.5a**), ~50–70% for OH(A–X) (**FIG. 5b**), almost triples for $N_2$(C–B) (**Fig. 5c**) and roughly doubles for $N_2$(B–A) (**Fig. 5d**). For the Ar I lines (772 nm in **Fig.5e** and 696.5 nm in **Fig.5f**) the effect is more evident, about double increase at the peak, although they have less intense peak intensities than the rest. A 50 V bias voltage produces little change for the molecular bands but already yields a noticeable rise for the Ar lines, while bias of 80–100 V uniformly boosts all emissions. Taken all together, these data indicate that the voltage bias strengthens gas excitation in the low pressure chamber which is most efficient for electron-impact-driven channels (Ar lines and $N_2$(C–B)).

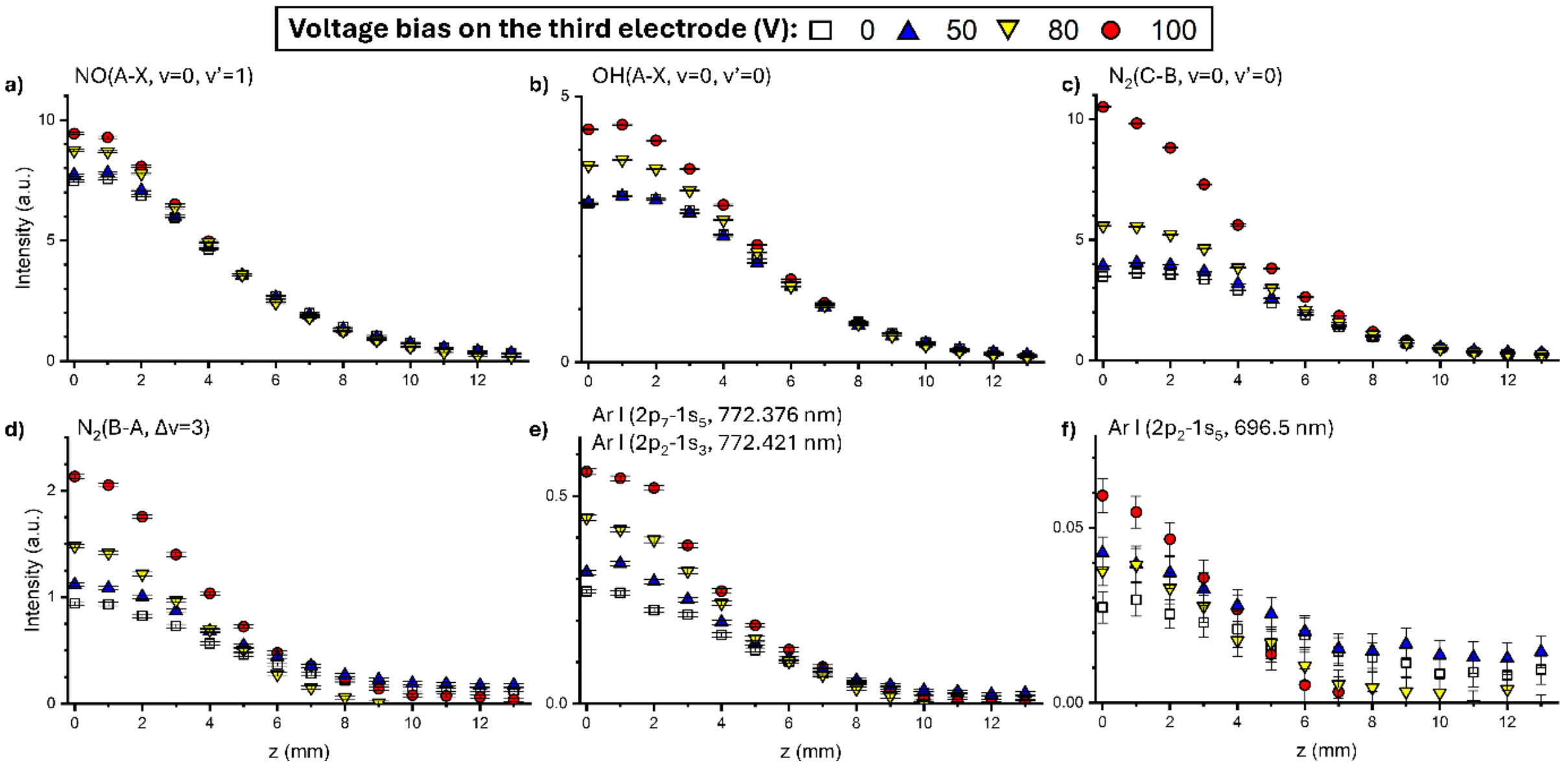

**Fig. 5**. Spatially resolved intensities of representative emissive species along the jet axis versus the voltage bias of the third electrode for a 50%–50% Ar/$N_2$ mixture, $P_{high}$=30 mbar, $P_{low}$=10 mbar and $I_d$=2.5 mA: **a**) NO(A–X, Δv=-1), **b**) OH(A–X, Δv=0), **c**) $N_2$(C–B, Δv=0), **d**) $N_2$(B–A, Δv=3), **e**) Ar I ($2p_7$–$1s_5$ and $2p_2$–$1s_3$), **f**) Ar I ($2p_2$–$1s_5$). These measurements are conducted in triplicate, with the resulting error bars representing these repetitions.

**Fig.6** shows the relative emission intensities of the aforementioned excited species (at z=0 mm) versus the discharge current (x-axis) for different Ar fractions in the Ar/$N_2$ mixture ($P_{high}$ = 30 mbar, $P_{low}$ = 10 mbar). Different colors/symbols shown represent different argon percentages (from 40%Ar, yellow, up to 100%Ar, red). Both the argon content and the discharge current generally affect the emission intensities of species generated. At z=0 mm, all emissions in **Fig.6** strengthen with current, but each species shows a distinct dependence on the Ar/$N_2$ ratio that reflects its formation and quenching pathways. In **Fig.6a**, NO(A–X) rises almost monotonically with current and is strongly favored in $N_2$-richer mixtures (40–60 % Ar), reaching intensities an order of magnitude above those in pure Ar at ≥3 mA. This reveals a greater availability of N– and/or O–atoms precursors with increasing $N_2$ percentage compared to operation in pure Ar. In **Fig.6b**, OH(A–X) also increases with current for every mixture, but the gas composition shows a smaller effect. A mild threshold/steepening appears in pure Ar near ~1–1.5 mA, while in 90% Ar a broad maximum emerges around ~2–2.5 mA, indicating that higher currents favor water dissociation and greater production of OH radicals.

In **Fig.6c**, the variation of the $N_2$(C–B) band emission is shown. For 100% Ar one can identify an inflection around 1 mA, beyond which the slope temporarily flattens (e.g. only a 10% increase from 0.75 to 1.25 mA, suggesting an early saturation). Past ~1.5 mA the intensity begins rising again more steeply, almost doubling between 1.5 and 3 mA. This two-stage behavior (a plateau followed by renewed growth) hints at changing excitation dynamics, possibly a critical Ar metastable production that allows more $N_2$ to be excited. In contrast, at 90% Ar (with only 10% $N_2$ in the mixture), the $N_2$(C–B) intensity shows a maximum around 2–2.5 mA, which is the highest intensity recorded for any species. Then, it declines by ~20% at 3.5 mA. This indicates that in a 90% Ar/10% $N_2$ mixture, there is an optimal current (~2–2.5 mA) for exciting $N_2$(C) beyond which further increases in current reduce the emission, likely due to overheating or quenching (e.g. enhanced collisional de-excitations at higher gas temperature, or a shift in discharge structure). Mixtures with lower Ar fractions (50–80% Ar) do not exhibit a sharp peak within the 0–3.5 mA range. Instead, $N_2$(C–B) tends to increase monotonically with current. In **Fig.6d**, the $N_2$(B–A) follows the same qualitative trend but more smoothly. In **Fig.6e,f**, the Ar I lines (772 nm and 696.5 nm) are more intense in pure Ar and are sharply suppressed by even modest $N_2$ additions; their growth with current becomes weak or saturating in $N_2$-rich mixtures, pointing to efficient quenching of Ar(2p) states and redistribution of electron energy into molecular channels. This agrees with the shrunk and distorted cathodic expansion observed in **Fig.2** for %$N_2$>10% in the mixture. Altogether, these behaviors are consistent with a balance between metastable-assisted excitation and collisional quenching/energy partitioning (stronger in $N_2$-rich or hotter plasmas), which respectively explain the 90% Ar optimum for $N_2$(C–B), the composition-insensitive yet current-driven OH response, the NO preference for $N_2$-rich feeds, and the suppression of Ar lines as $N_2$ percentage increases. A more detailed discussion on the formation mechanisms of key species is done in the next section.

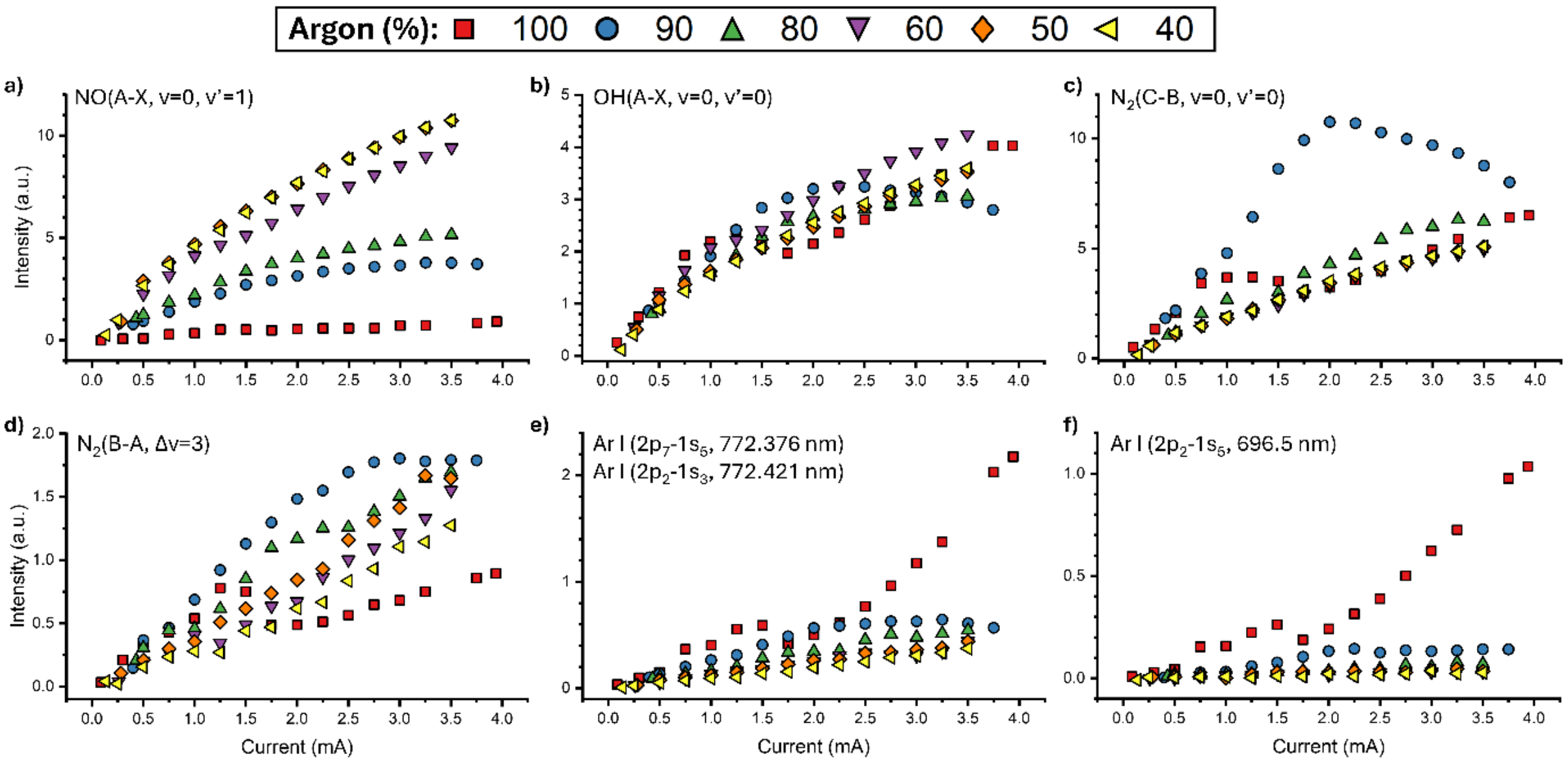

**Fig. 6**. Emission intensities of various excited species (recorded at z=0 mm and using unbiased third electrode) versus the discharge current for different argon fractions in the Ar/$N_2$ mixture: **a**) NO(A–X, Δv=-1), **b**) OH(A–X, Δv=0), **c**) $N_2$(C–B, Δv=0), **d**) $N_2$ (B–A, Δv=3), **e**) Ar I ($2p_7$-$1s_5$ and $2p_2$-$1s_3$), **f**) Ar I ($2p_2$-$1s_5$). $P_{high}$ = 30 mbar, $P_{low}$ = 10 mbar.

## Rotational Temperature determination from molecular spectra

### 1. $N_2$ (C-B) bands

The $N_2$(C-B) emission band is one of the most used bands to estimate $T_{gas}$ from the rotational temperature of the $N_2$(C) state. In our conditions, this is one of the most prominent emissions observed in the spectra of the MHCD (**Figs. 3, 4**). It is still observable even for 100% Ar due to some residual impurities in the reactor. The formation of the $N_2$(C) state can be induced by direct electron impact from $N_2$(X), $N_2$(A) and $N_2$(B) states, as follows:

e + $N_2$(X) → e + $N_2$(C), k=1.3 × $10^{-19}$–2.7 × $10^{-16}$ $m^3s^{-1}$, $T_e$ = 1–3 eV (**R1**) [19]

e + $N_2$(A) → e + $N_2$(C), k=5.6 × $10^{-16}$–2.7 × $10^{-14}$ $m^3s^{-1}$, $T_e$ = 1–3 eV (**R2**) [19]

e + $N_2$(B) → e + $N_2$(C), k≈$10^{-15}$ $m^3s^{-1}$, $T_e$ =1 eV (**R3**) [33]

Other production mechanisms are likely responsible for the population of the $N_2$(C) state in the MHCD jet, like energy transfers from Ar($1s_3$)/Ar($1s_5$) metastables as well as $N_2$(A) pooling reactions:

$N_2$(A) + $N_2$(A) → $N_2$(C) + $N_2$(X), k=1.5 × $10^{-16}$ $m^3s^{-1}$ (**R4**) [19, 21]

$Ar^m$ + $N_2$ (X) → $N_2$(C) + Ar, k = 3.6 ×$10^{-17}$ $m^3s^{-1}$ (**R5**) [19, 21, 34]

The energy of the $N_2$(C) state is 11.03 eV, while those of Ar($1s_3$)/Ar($1s_5$) and $N_2$(A) are 11.55/11.72 and 6.17 eV, respectively. Thus, the above reactions can efficiently lead to $N_2$(C) excitation to different ro-vibrational levels, resulting in the rotational population distribution to not be in equilibrium with the $T_{gas}$ [19, 21]. The formation of $Ar^m$ and $N_2$(A) under our conditions is witnessed by the various radiative transitions shown in the emission spectrum of **Fig. 3** (see, e.g., $N_2$(FPS) bands between 630 and 690 nm, and Ar I lines at 763.5, 772.4, 794.8 and 811.5 nm). Furthermore, the density of $Ar^m$ has been directly measured in a previous study on the same MHCD [17]. Thus, it is likely that the rotational distribution of the $N_2$(C) state is described by different rotational temperatures, representative of the different excitation pathways of the excited state involving both electrons and metastables.

**Fig**. **7a** shows $N_2$(C-B, Δv=0) spectra with band heads at 337.14 nm (see **Fig.4** for detailed identification) for selected operating conditions of the MHCD: various Ar-$N_2$ mixtures (10%-90%, 50%-50%, 90%-10% and 100%-0%) at $P_{high}$=80 mbar, $P_{low}$=10 mbar and $I_d$=2.5 mA. As can be seen in the inset, at high Ar percentages in the mixture, numerous rotational lines are developed in the range 333–335 nm corresponding to a range of high rotational quantum numbers (see also **Fig.4**). These lines are barely seen at low Ar percentages (below 10%Ar) indicating an important role of Ar content on the rotational population of the $N_2$(C) level. Indeed, the experimentally measured $T_{rot}$ (here obtained by performing a fit with a single $T_{rot}$) increases from 564 K to 1542 K when changing the %Ar in the mixture from 10%Ar to 100%Ar. The main reaction responsible for this increase in $T_{rot}$ is (**R5**). This result reveals that a single temperature fit only returns an "average $T_{rot}$" between a smaller and a larger value corresponding to different excitation pathways. In the former case, $N_2$(C) excitation is dominated by electrons (reactions (**R1**-**R3**)) and the rotational distribution population of the ground state is maintained at the excited state as well [33, 35]. Thus, the rotational temperature of the upper state is a good estimate of that of the ground state (which is in equilibrium with the translational temperature, i.e., the $T_{Gas}$). In the latter case, however, excitation induced by $Ar^m$ or $N_2$(A) leads to an overpopulation of the rotational distribution of the excited state and thus its $T_{rot}$ deviates from that of the ground state. In state-resolved afterglow experiments, Nguyen and Sadeghi [34] depopulated Ar($1s_3$) and Ar($1s_5$) states by optical pumping and monitored $N_2$(C-B) radiation. They explicitly showed a two-component rotational distribution of $N_2$(C, v′=0), composed of a "cold" group and a "hot" tail formed because of the $Ar^m$ energy transfer dynamics. This explains why, in our conditions, increasing the Ar content (hence the $Ar^m$ density) in the $N_2$/Ar mixture systematically increases the average $N_2$(C) rotational temperature value obtained with a single temperature fitting. Methodologically, when $N_2$(C) excitation is dominated by $Ar^m$ energy transfers, $T_{rot}$ of $N_2$(C) should be interpreted as a $T_{Gas}$ overestimation. Thus, good practices to infer $T_{Gas}$ in such conditions are to perform a two-temperature fit of the ro-vibrational band, or cross-validate against the $T_{rot}$ of other probe species which are less sensitive to such disturbances. In the present work, we particularly aim to perform a cross-validation of $T_{rot}$ of different molecules formed in the discharge.

Another notable result is the formation of the NH($A^3\Pi$-$X^3\Sigma^-$, Δv=0) triplet system around 336 nm which, under certain conditions, overlaps with the $N_2$(C-B) band. Specifically, this emission band is prevalent when the gas mixture becomes rich in nitrogen, and its intensity increases with decreasing $P_{high}$ (**Fig.7b**). This band strongly distorts the actual $N_2$(SPS) emission spectrum and should be deconvoluted from the raw experimental profile before performing a fitting of the rotational band and extracting any $T_{rot}$. If this is not considered, it may also lead to an overestimation of the rotational temperature in our case.

Ground state NH(X) can form through reactions involving $H_2O$ impurities and N atoms, as follows:

$$N(^2D/^2P) + H_2O \rightarrow NH(X) + OH, \; k = 2.4\times10^{-18} \; m^3s^{-1} \quad \textbf{(R6)} \; [36–38]$$

NH(A) formation from NH(X) requires a relatively small energy (3.72 eV) which can be induced by electron impacts. $H_2O$ is coming as an impurity from the gas bottles (see **Table 1**) or from potential humidity residuals in the chamber walls. Besides, this MHCD can efficiently dissociate $N_2$ and produce high densities of N atoms (between $10^{13}$ and $10^{15}$ $cm^{-3}$ depending on the operating conditions) as it has been demonstrated in previous works [6, 17]. This makes (**R6**) relevant under our conditions.

The production of N atoms can occur through pooling reactions between $N_2$(A) metastables, between $Ar^m$ and $N_2$ molecules, even by dissociative recombination of $N_2^+$ (which is identified in our spectrum; see **Fig.3**) [17]. These are described with the following reactions [19]:

$$N_2(A) + N_2(A) \rightarrow N_2 + N + N, \; k = 3\times10^{-17} \; m^3s^{-1} \; m^3s^{-1} \quad \textbf{(R7)}$$

$$Ar^m + N_2(A/B/C, v) \rightarrow N + N + Ar, \; k = 3.6\times10^{-17} \; m^3s^{-1} \quad \textbf{(R8)}$$

$$e + N_2^+ \rightarrow N + N, \; k = 4.8\times10^{-13}\times(T_e(K)/300)^{0.5} \; m^3s^{-1} \quad \textbf{(R9)}$$

Although a detailed analysis of the NH(A-X) emission is not a major focus in the present study, its overlap with $N_2$(C-B) bands highlights the need to account for the impact on $T_{rot}$ of potential overlapping emissions from minor species. To mitigate this issue in our conditions, a deconvolution procedure is applied to isolate the NH(A-X) and $N_2$(C-B) bands. To this end, the Moose Python library has been utilized [30], which is based on the MassiveOES [24–26] software allowing to simultaneously fit multiple species (even overlapping ones) by concatenating their line strength files. The convoluted fitting of NH(A-X) and $N_2$(C-B) transitions is demonstrated in **Fig.8a**, for a case where the NH(A-X) intensity is clearly distinguishable ($P_{high}$=20 mbar, $P_{low}$=10 mbar, $I_d$=2.5 mA current, 50% Ar – 50% $N_2$), while the individual deconvoluted fits are shown in **Fig.8b**.

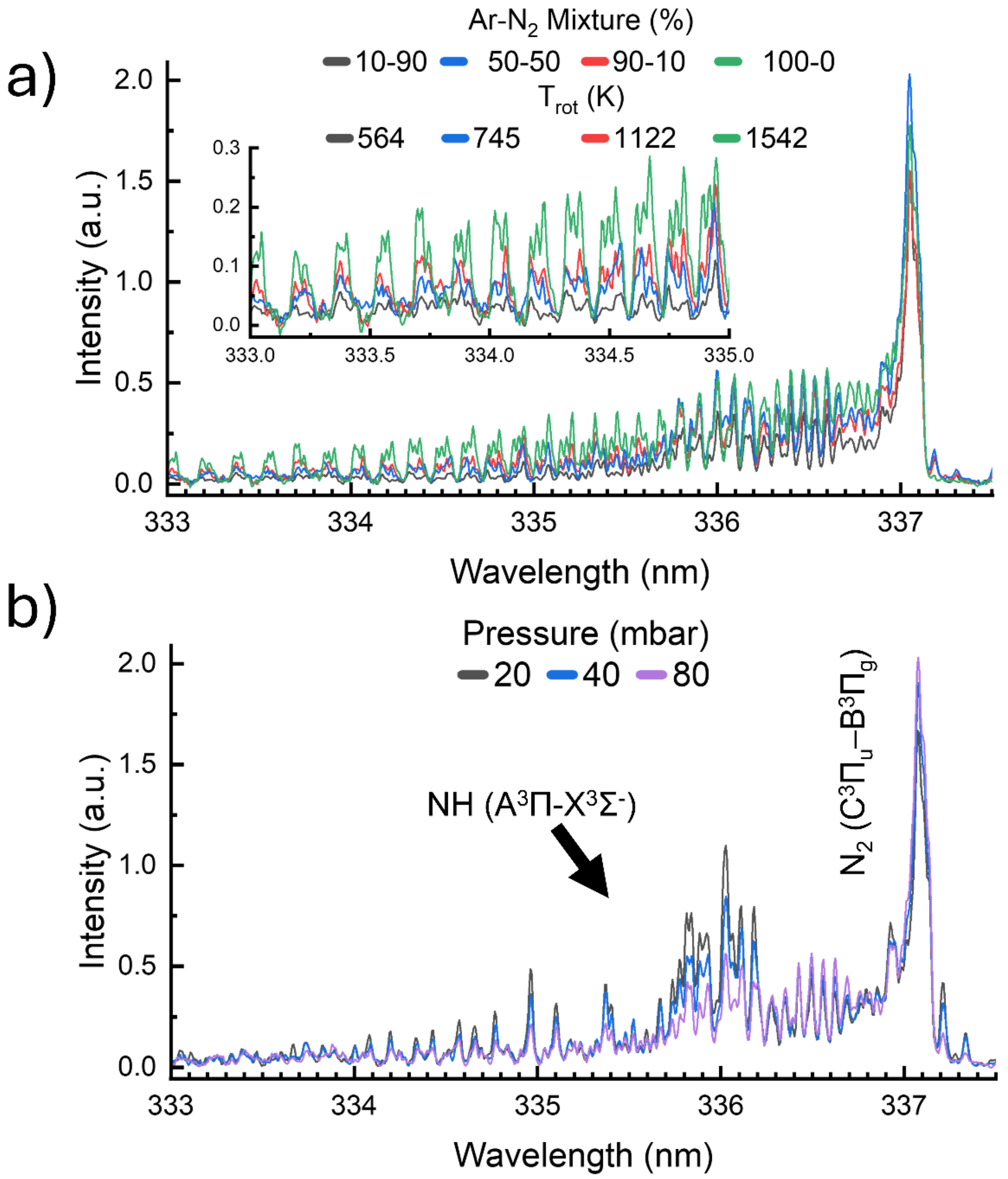

**Fig.7**. **a**) $N_2$(C-B, Δv=0) emissions for different Ar-$N_2$ mixtures at $P_{high}$=80 mbar, $P_{low}$=10 mbar and $I_d$=2.5 mA. **b**) $N_2$(C-B, Δv=0) and NH(A-X, Δv=0) emissions for different $P_{high}$ at an Ar-$N_2$ mixture of 50%-50% and $I_d$=2.5 mA.

Based on this procedure, the $T_{rot}$ of both excited states can be obtained (**Fig. 8**). **Fig. 8c** shows the variation of $T_{rot}$ of $N_2$(C) and NH(A) versus the discharge current at $P_{high}$=30 mbar and 50%-50% Ar-$N_2$ mixture. If NH(A-X) emission is not considered on the fitting, the average $T_{rot}$ of the raw spectrum can easily be overestimated to be around 1000 K or more, while it does not show any variation with the current. Instead, after deconvolution and independent fitting of each band, the $T_{rot}$ of $N_2$(C) is found to be about two times smaller and shows a dependence on the current (ranging between 450 and 550 K for 0.5 and 3 mA, respectively). Besides, the $T_{rot}$ of NH(A) is found to vary between 280 and 360 K for the same current values, being slightly smaller than that of $N_2$(A). This finding highlights the importance of performing a critical analysis of rovibrational emissions in cases where unexpected (or unpredicted) molecular emissions are formed in the discharge. Other strong $N_2$(SPS) emissions are identified in **Fig. 3**, e.g., around 357 nm and 380 nm. These are not affected by any other emission and are also used for $T_{rot}$ estimation of $N_2$(C) for cross-comparison and more robust analysis.

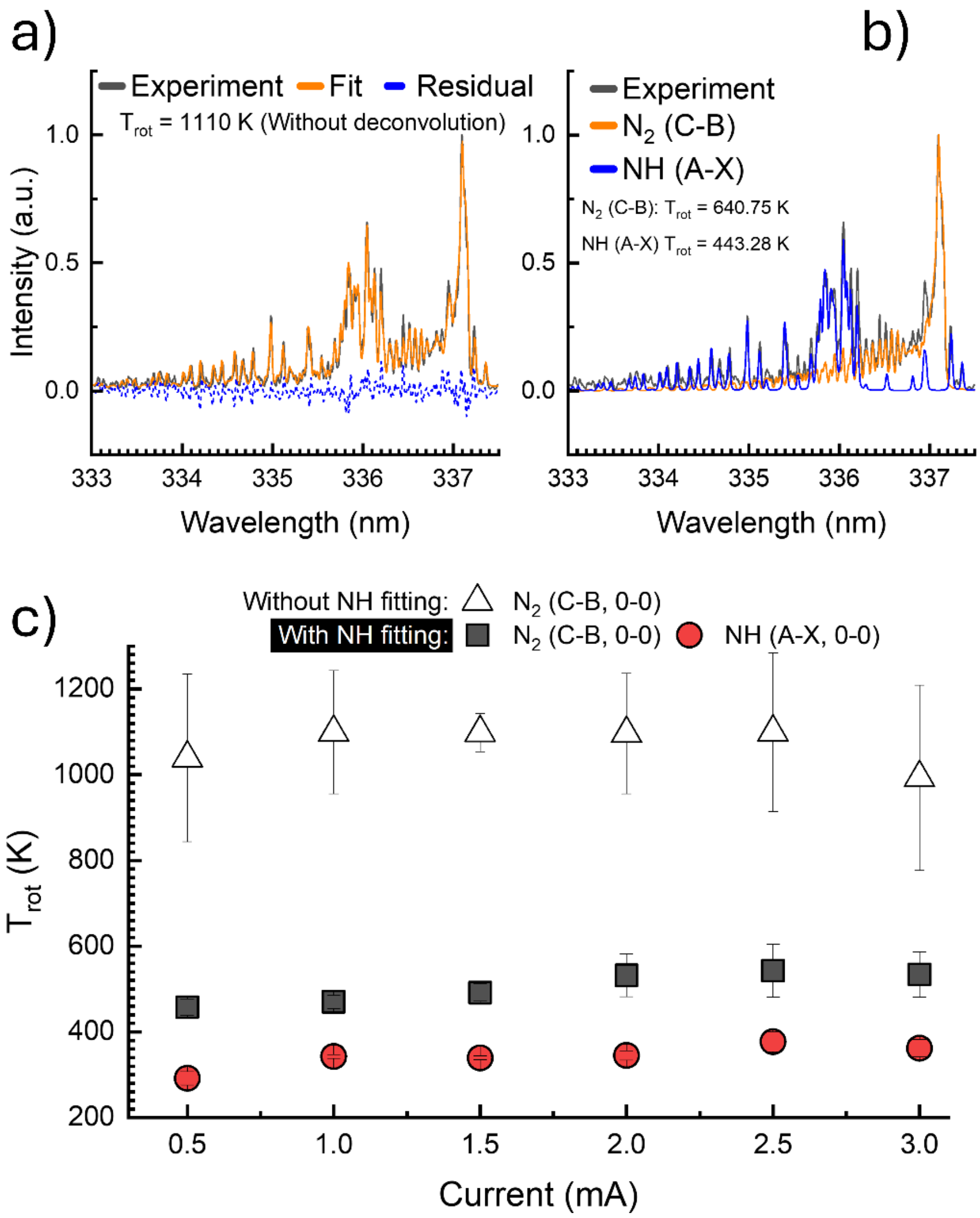

**Fig.8**. **a**) Fitting examples of the experimental spectra with synthetic spectra (through Moose/MassiveOES) and simultaneous determination of $T_{rot}$ of $N_2$(C) and NH(A) at $P_{high}$=20 mbar, $P_{low}$=10 mbar, $I_d$=2.5 mA. **b**) Deconvolution of NH(A-X) and $N_2$(C-B) emissions from the total experimental profile and determination of $T_{rot}$ of NH(A) and $N_2$(C). The error values come from the fitting uncertainty. **c**) Determined $T_{rot}$ values of $N_2$(C) and NH(A) versus $I_d$ ($P_{high}$=30 mbar, $P_{low}$=10 mbar, 50%-50% Ar-$N_2$). These measurements were conducted in triplicate, with the resulting error bars representing these repetitions.

**Fig.9** summarizes the determination of the vibrational temperature of $N_2$(C–B) in the MHCD jet. In **Fig.9a** the experimentally recorded SPS (Δv=–2) band is well reproduced by a synthetic spectrum generated with MassiveOES, yielding an average vibrational temperature of $T_{vib,mean}$=4480±120K, which confirms that the observed band system can be described by a single $T_{vib}$. This is further validated in **Fig.9b**, where the Boltzmann plot of the vibrational populations (manual data treatment/plot) exhibits a clear linear behavior within the error margin; the slope of the linear fit gives a similar value of $T_{vib,mean}$=4400±150K, showing the internal consistency of the two approaches. The evolution of $T_{vib}$ with the discharge current is shown in **Fig.9c** for two representative gas mixtures (100% Ar and 50% Ar – 50% $N_2$) at $P_{high}$=30 mbar and $P_{low}$=10 mbar. In the case of 100%Ar $T_{vib}$ increases slightly with current, possibly due to an enhanced production of $Ar^m$ and subsequent $N_2$(C) excitation. Besides, the mixed Ar/$N_2$ plasma systematically exhibits larger vibrational temperatures (up to ~5000 K) compared to pure Ar, indicating that the presence of a decent amount of $N_2$ facilitates vibrational excitation of nitrogen. In this condition, an insignificant variation with the current is revealed. This could be due to the fact that the reactions leading to the formation of the excited state have reached a certain state of balance being governed by Ar metastables and the excitation of $N_2$(C) is no more affected by an increase of the current. Comparing these $T_{vib}$ with the much lower corresponding $T_{rot}$ found in **Fig.8** shows a significant difference, which is typical of a strongly non-equilibrium plasma. This confirms that the discharge operates far from thermal equilibrium.

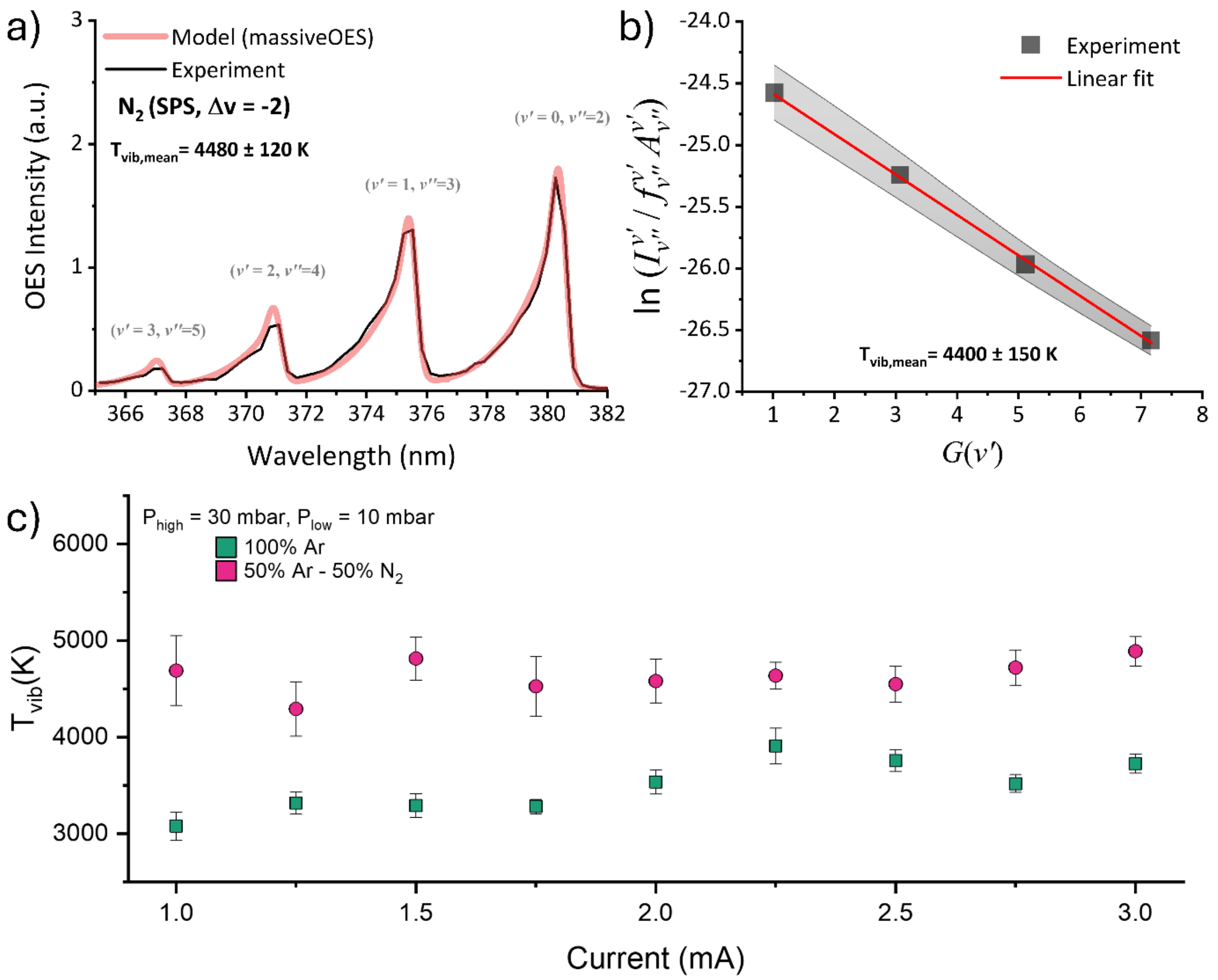

**Fig.9**. Determination of $T_{vib}$ of $N_2$(C) via **a**) fitting of the experimental spectrum with a synthetic spectrum referring to a single $T_{vib}$, and its **b**) corresponding Boltzmann plot for the $N_2$(C-B) vibrational distribution (Δv=–2, $P_{high}$=30 mbar, $P_{low}$=10 mbar, $I_d$=2.5 mA, 50% Ar – 50% $N_2$). **c**) $T_{vib}$ of $N_2$(C) (obtained through the Boltzmann plot) against $I_d$ for different Ar/$N_2$ mixtures. The error bars originate from the fitting uncertainty.

**OH (A-X, Δv=0) band**

The OH($A^2\Sigma^+$–$X^2\Pi_i$) emission with characteristic rotational spectrum around 309 nm is also widely used to infer the $T_{Gas}$ in non-equilibrium plasmas [19, 21, 26, 39]. Fortunately, this molecule is also formed under our experimental conditions. Hydroxyl formation originates in water impurities present in the gas bottles (see Table 1) or accumulated in the reactor walls, as already mentioned. Specifically, OH(X) can be produced by electron dissociation, as follows [19]:

e + $H_2O$ → OH(X) + H + e, k = (2.3×$10^{-18}$ - 8.6×$10^{-16}$) $m^3s^{-1}$ for $T_e$ = (1-3) eV (**R10**)

Then, the OH(A) state can be populated through electron impact excitation [19]:

e + OH(X) → e + OH(A), k = (3.2×$10^{-14}$ - 7.5×$10^{-13}$) $m^3s^{-1}$ for $T_e$ = (1-3) eV (**R11**)

, or be populated directly through electron-induced water dissociative excitation [19]:

e + $H_2O$ → OH(A) + H + e, k = (1.6×$10^{-19}$ - 1.2×$10^{-16}$) $m^3s^{-1}$ for $T_e$ = (1-3) eV (**R12**)

However, other mechanisms of OH and OH(A) formation can be driven by $Ar^m$ and $N_2$(A) metastables [19]:

$N_2$(A) + OH(X) → $N_2$(X) + OH(A), k = $10^{-17}$ $m^3s^{-1}$ for $T_{Gas}$ = 300 K (**R13**)

$Ar^m$ + $H_2O$ → OH + H +Ar, k = 4.5×$10^{-16}$ $m^3s^{-1}$ for $T_{Gas}$ = 300 K (**R14**)

In the present MHCD, a part of the OH(A-X) transition detected between 306 and 312 nm is perturbed from the $N_2$(C-B, Δv=1) band between 310 and 315 nm, as can be seen in **Fig.10a**. Fitting the distorted raw OH(A-X) spectrum may give rise to uncertainties on the corresponding $T_{rot}$. Thus, when the OH(A-X) spectrum is distorted from the $N_2$(SPS) band, only the experimental spectrum contained in the spectral region 306–311 nm is used in the fitting procedure. This is still acceptable as has been also used in other works as well [19]. For 100%Ar and small $N_2$ contents (<20%) in the mixture, the OH(A-X) spectrum is not affected by the $N_2$(SPS), and the full spectrum can be used for $T_{rot}$ analysis.

Furthermore, as in the case of the $N_2$(C), the relative contribution of reactions (**R10**)-(**R14**) to the population of OH(A) affects its $T_{rot}$. **Fig.10b** shows an example of the determined $T_{rot}$ of OH(A) by fitting a synthetic spectrum (black line) on an experimentally recorded spectrum (red line). Despite avoiding disturbance from the adjacent $N_2$(SPS) spectrum, the obtained $T_{rot}$ still can be affected by the contribution of metastables, resulting in an average $T_{rot}$ of about 620 K. Indeed, using the corresponding Boltzmann plot (**Fig.10c**), two groups (yellow and green points) of rotational populations can be distinguished yielding two temperatures which differ between them by almost a factor of 8. These are indicative of 'cold' (yellow points) and "hot" groups of rotationally excited molecules. The linear fitting of the "cold" group reveals an average $T_{rot}$ of about 530 K, which is about 85 K smaller than that obtained in **Fig.10a**. Its population is expected to be due to reactions (**R11**) and (**R12**) while that of the "hot" group is driven by reactions (**R13**) and (**R14**). Consequently, the $T_{rot,OH}$ of the "cold" group is most likely in equilibrium with that of the ground state which is expected close to that of the gas [19, 39].

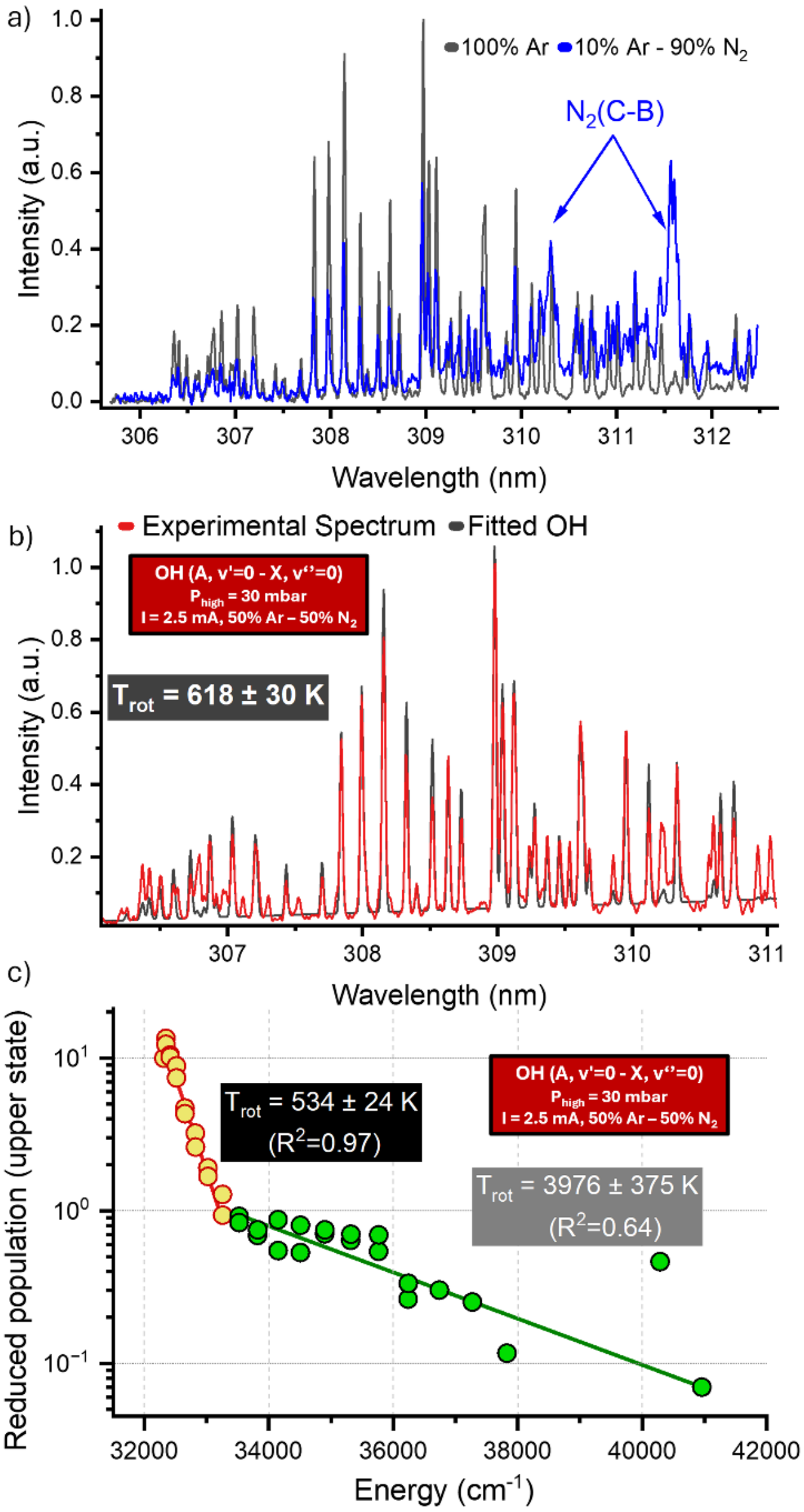

**Fig. 10**. **a**) OH(A-X) emission spectra at 100%Ar and 10%Ar – 90%$N_2$ ($P_{high}$=30 mbar, $P_{low}$=10 mbar, $I_d$=2.5 mA). Determination of $T_{rot}$ of OH(A) via **b**) fitting of the experimental spectrum with a synthetic spectrum referring to a single

$T_{rot}$, and **c**) corresponding Boltzmann plot for the OH(A-X) rotational distribution ($P_{high}$=30 mbar, $P_{low}$=10 mbar, $I_d$=2.5 mA, 50% Ar – 50% $N_2$).

**Fig.11a-c** show the $T_{rot}$ values obtained from simulating the OH(A-X) spectrum (assuming a single $T_{rot}$), under various currents, pressures and Ar/$N_2$ mixtures. **Fig.11a** shows the variation of $T_{rot}$ for different discharge currents and pressures in 50%-50% Ar-$N_2$ mixture. $T_{rot}$ spans from around 350 K at the lowest discharge currents, where the discharge ignites, up to 600 K for higher currents. The inset in **Fig.11a** compares the $T_{rot}$ values obtained with a single-temperature fit with those obtained from the Boltzmann plot method at 30 mbar for 1 mA and 2.5 mA. Notably, in these discharge conditions, the smallest and largest differences between the methods were observed for these two values, respectively. An increase of the discharge current yields higher deposited power in the MHCD. This influences the ro-vibrational excitation of the probe molecules studied. Furthermore, it seems that higher pressures ($P_{high}$) lead to lower rotational temperatures. In all cases, increasing the current leads to a steady rise in $T_{rot}$. Besides, the effect of the gas composition on $T_{rot}$ is noticeable in the contour plots of **Fig.11b** and **Fig.11c**. These also show the variation of $T_{rot}$ for different applied voltages to the third electrode and different positions on the axis of the jet. As can be seen, mixtures richer in nitrogen (**Fig.11b**) exhibit higher OH(A) $T_{rot}$ values than mixtures poorer in nitrogen (**Fig.11c**), probably due to metastable-driven excited state population (**R13** and **R14**). Overall, from **Fig.11b** and **Fig.11c**, it is evident that the voltage bias of the third electrode (in the range of values studied here) has no significant effect on the $T_{rot}$ deduced from OH(A). The OH(A) $T_{rot}$ also remains relatively constant along the jet axis.

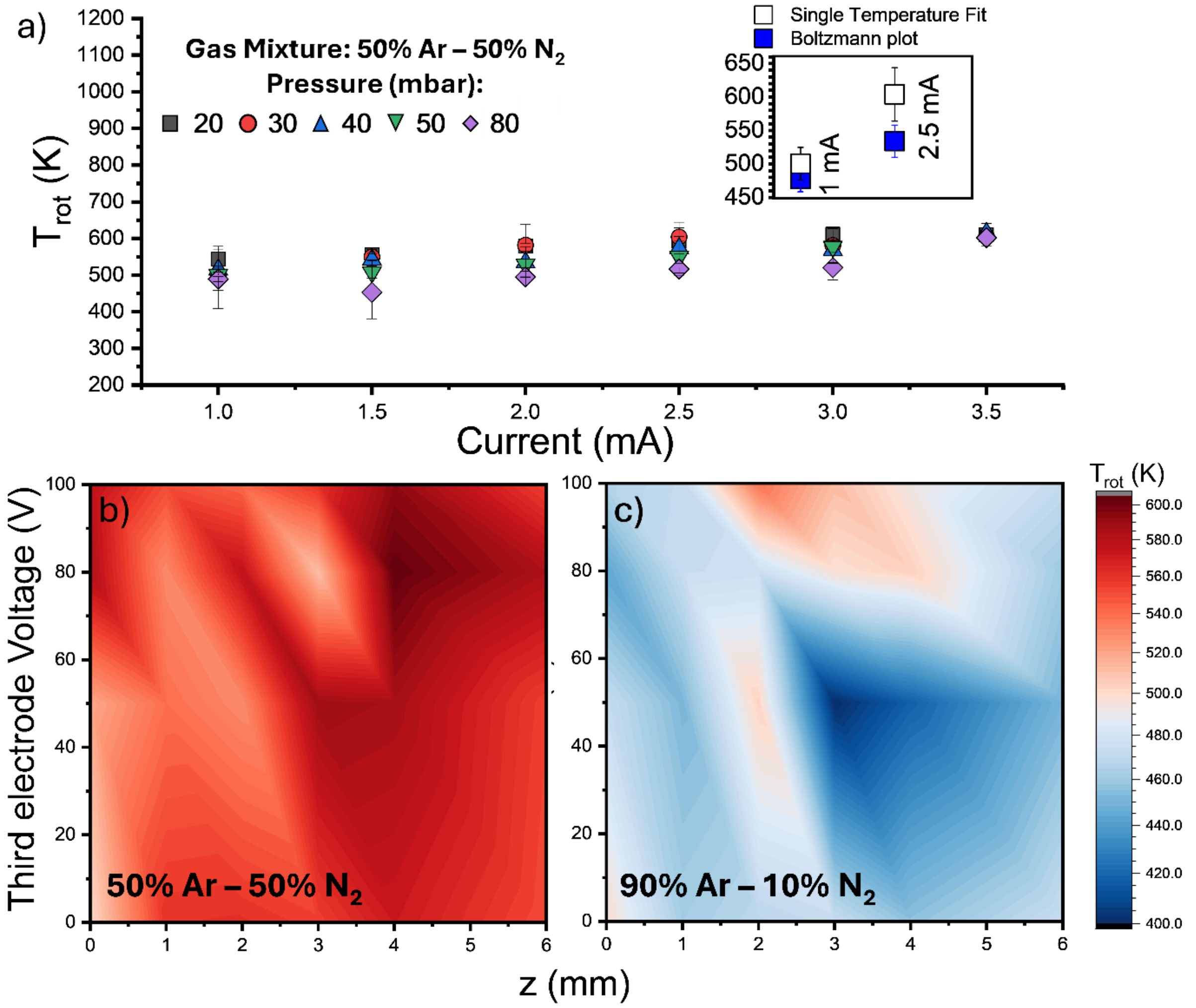

**Fig. 11**. **a**) $T_{rot}$ of OH(A) (obtained through single $T_{rot}$ fitting, as shown in **Fig.10a**) against $I_d$ for different $P_{high}$. These measurements were conducted in triplicate, with the resulting error bars representing these repetitions. The inset shows the comparison of $T_{rot}$ values obtained through single-temperature fitting and the Boltzmann plot method (as shown in **Fig.10b**) for two indicative values of $I_d$. Distribution of $T_{rot}$ of OH(A) along the jet axis for different applied voltages to the third electrode at $P_{high}$=30 mbar for gas mixtures of **b**) 50% Ar – 50% $N_2$ and **c**) 90% Ar – 10% $N_2$.

**NO (A-X, Δv=-1) band**

In the present conditions, the NO($A^2\Sigma^+$–$X^2\Pi$) emission, also known as the NOγ system can be observed as well. Ground state NO(X) can be formed due to the presence of trace OH-containing impurities (e.g., from water vapor) reacting with N atoms (**R15**) produced by $N_2$ dissociation (**R7-R9**), or by reaction of O-atoms with the $N_2$(A) metastable state (**R16**).

$$N + OH \rightarrow NO + H, \quad k \approx 4\exp(85/T_{Gas}) \times 10^{-17}\ m^3 s^{-1}\ (\mathbf{R15})\ [21, 40]$$

$$N_2(A) + O \rightarrow NO + N(^2D), \quad k = 7 \times 10^{-12}\ cm^3\ s^{-1}\ (\mathbf{R16})\ [41]$$

The NO(A) state can be populated through electron-impact excitation of NO(X) (**R17**) but may also result from energy transfer reactions involving metastable $N_2$(A) molecules (**R18**).

$$NO(X) + e \rightarrow NO(A), \quad k=f(E/N)\ (\mathbf{R17})\ [42, 43]$$

$$N_2(A)+NO(X) \rightarrow NO(A)+N_2(X), \quad k=(6.5\text{–}7.8)\times 10^{-11}\ cm^3\ s^{-1}\ (\mathbf{R18})\ [21, 41\text{–}43]$$

**Fig.12a** shows an example of the NO(A–X, $\Delta v = -1$) emission spectrum recorded at the jet axis (z=0 mm) for a 90% Ar – 10% $N_2$ mixture at $P_{high}$=30 mbar and $I_d$ = 2.5 mA. A synthetic spectrum generated using MassiveOES (black line) from ExoMol data [31] was fitted to the experimental data (red line), yielding a single rotational temperature of 850±30 K, which is a relatively high value. Nevertheless, the potential influence of non-equilibrium excitation channels cannot be excluded, and the inferred $T_{rot}$ may overestimate the true gas temperature, as noted previously. The spatial evolution of $T_{rot}$ of NO(A) on the jet axis is presented in **Fig.12b** for two gas mixtures: 90% Ar – 10% $N_2$ and 50% Ar – 50% $N_2$, both recorded at $I_d$=2.5 mA, $P_{high}$=30 mbar and $P_{low}$=10 mbar. In both cases, $T_{rot}$ remains nearly constant along the jet axis, within the experimental uncertainty. No significant drop or rise is observed up to z=10 mm from the MHCD hole, which suggests that the NO(A) rotational distribution is either quickly established near the hole or is sustained by similar excitation and relaxation conditions throughout the jet length. Higher $T_{rot}$ values are obtained for the Ar-rich mixture, potentially reflecting increased NO(A) production due to more effective energy transfer from $N_2$(A) metastables (**R18**). This is based on the fact that enhanced $N_2$(FPS) emission is obtained at 90%Ar-10%$N_2$, thus leading to a larger population of the $N_2$(A) states (**Fig.6**) which participate in (**R18**). The measured $T_{rot}$ values derived from NO(A) emission at 50%Ar - 50%$N_2$ in **Fig.12b** agree relatively well with the $T_{rot}$ derived from OH(A) and $N_2$(C). Furthermore, this aligns with multi-species investigations in dielectric barrier discharges [44], which confirm that NO can serve as a valuable complementary $T_{Gas}$ probe in non-equilibrium environments. However, caution must be exercised when interpreting NO rotational temperatures, as excitation by $N_2$(A) metastables may lead to non-thermal rotational distributions [21, 42]. Prior studies have shown that NO(A) rotational temperatures can overestimate $T_{Gas}$, especially when formed via nascent or metastable-driven mechanisms [19, 45]. Therefore, NO(A–X) emission may offer reasonable estimates of $T_{Gas}$ under certain operating conditions of the present study but should ideally be cross-validated with other thermometric species such as $N_2$ and OH.

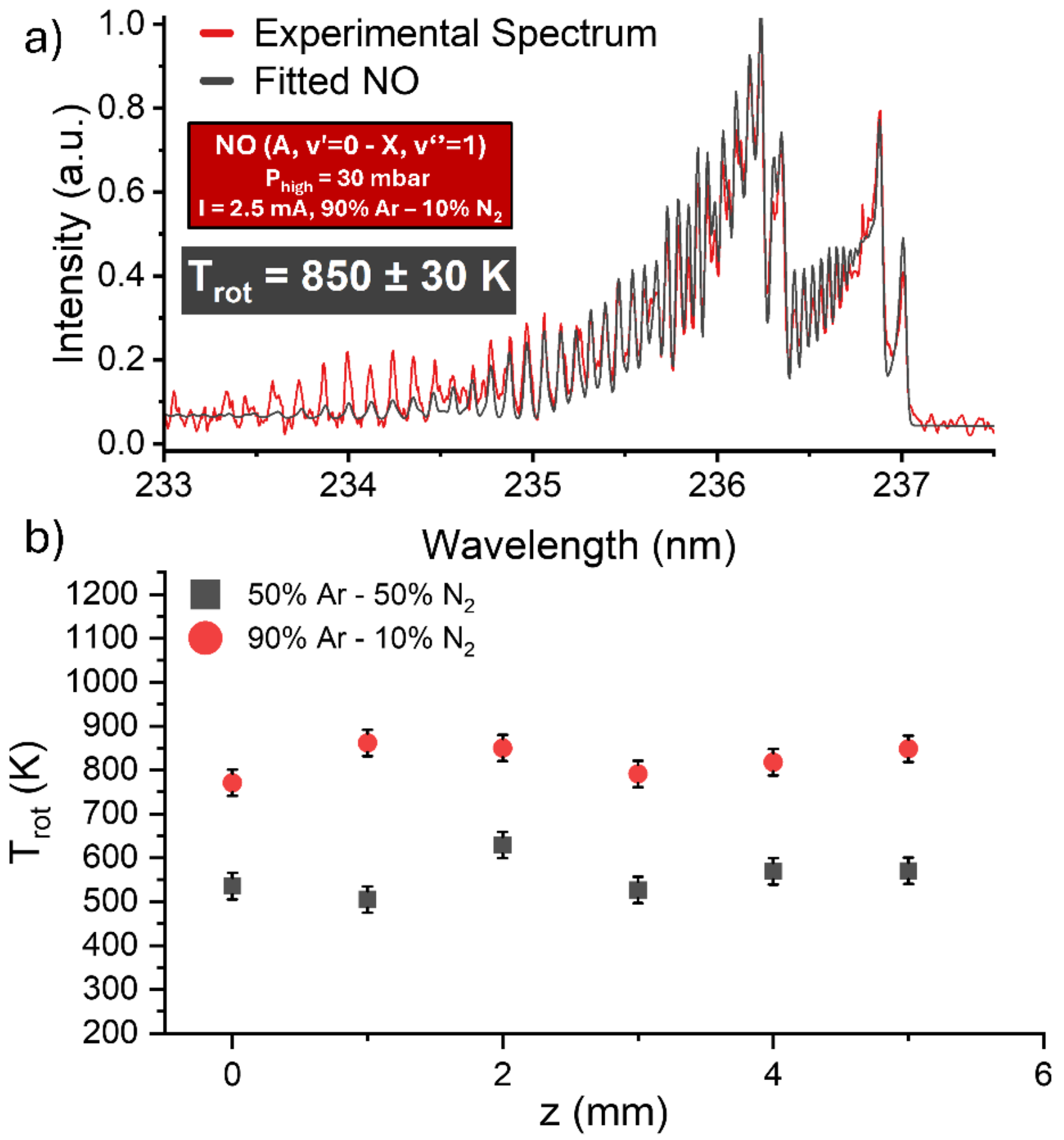

**Fig.12**. Determination of $T_{rot}$ of NO(A) via **a**) fitting of the experimental spectrum with a synthetic spectrum referring to a single $T_{rot}$. **b**) Distribution of $T_{rot}$ of NO(A) along the jet axis at $I_d$=2.5 mA, $P_{high}$=30 mbar, $P_{low}$=10 mbar for gas mixtures of 50% Ar – 50% $N_2$ and 90% Ar – 10% $N_2$. The error bars originate from the fitting uncertainty.

## Cross-comparison of $T_{rot}$ of different molecules

**Fig.13a** shows the comparison between the $T_{rot}$ values of OH(A) and $N_2$(C) as a function of the discharge current for $P_{high}$=30 mbar, $P_{low}$=10 mbar and a 50% Ar–50% $N_2$ gas mixture. The $T_{rot}$ of OH(A) are derived using a Boltzmann plot fitting procedure, while those of $N_2$(C) are determined by fitting and deconvolving the NH(A-X) system, as shown in **Fig.8**. For both molecules a consistent increase in $T_{rot}$ with the discharge current occurs, reflecting progressive gas heating in the discharge due to an increase of the deposited power, as expected. Importantly, the two molecules yield identical temperatures within the experimental uncertainty. This agreement suggests that, under the present operating conditions, the measured rotational temperatures of these molecules are likely representing the $T_{Gas}$ in the MHCD jet. Furthermore, **Fig.13b** shows the comparison of determined $T_{rot}$ of OH(A) (obtained using the single $T_{rot}$ fitting and Boltzmann plot methods), NO(A), NH(A) and $N_2$(C) (after deconvolution of NH(A-X) emission) obtained through fitting of the transitions around 237, 336 and 337 nm. The data points have an artificial z-offset applied to prevent overlap, which is solely for visualization. This is not a physical z-offset; the different z-measurements are accurately represented by the plot's axis ticks. These show the spatial evolution of $T_{rot}$ along the jet axis (z=0-5 mm) and highlight the impact of the gas mixture on the rotational temperatures extracted from the different molecules studied. In the case of the 50%Ar – 50%$N_2$ mixture (represented by squares), the temperatures derived from NH(A) and OH(A) are in relatively good agreement, varying between 450 K and 550 K along the jet, while the $N_2$(C) temperature is clearly higher (~650 K) implying an influence from Ar metastables. However, a significant divergence occurs in the Argon-rich mixture (90% Ar – 10% $N_2$, represented by circles). While the OH(A) temperature remains consistent with the values obtained in the balanced mixture (~500 K), the $T_{rot}$ for $N_2$(C) and NO(A) exhibit a sharp increase, reaching values exceeding 1100 K and 800 K, respectively. This discrepancy confirms that in Ar-rich conditions, the rotational populations of $N_2$(C) and NO(A) are strongly driven by non-thermal energy transfers from $Ar^m$ and/or $N_2$(A), leading to an overestimation of the gas temperature. Although NO(A) is not directly populated by $Ar^m$, its excitation can be dominated by energy transfer from $N_2$(A) metastables though (**R18**). In argon-rich conditions, the $N_2$(A) population may act as an energy reservoir that is fed by the radiative cascade of $N_2$(C), which is efficiently excited by collisions with $Ar^m$ through (**R5**). Conversely, the stability of OH(A) and NH(A) temperatures across different mixtures suggests these species are less sensitive to metastable pumping, making them the most reliable indicators of the true $T_{Gas}$ in this plasma source.

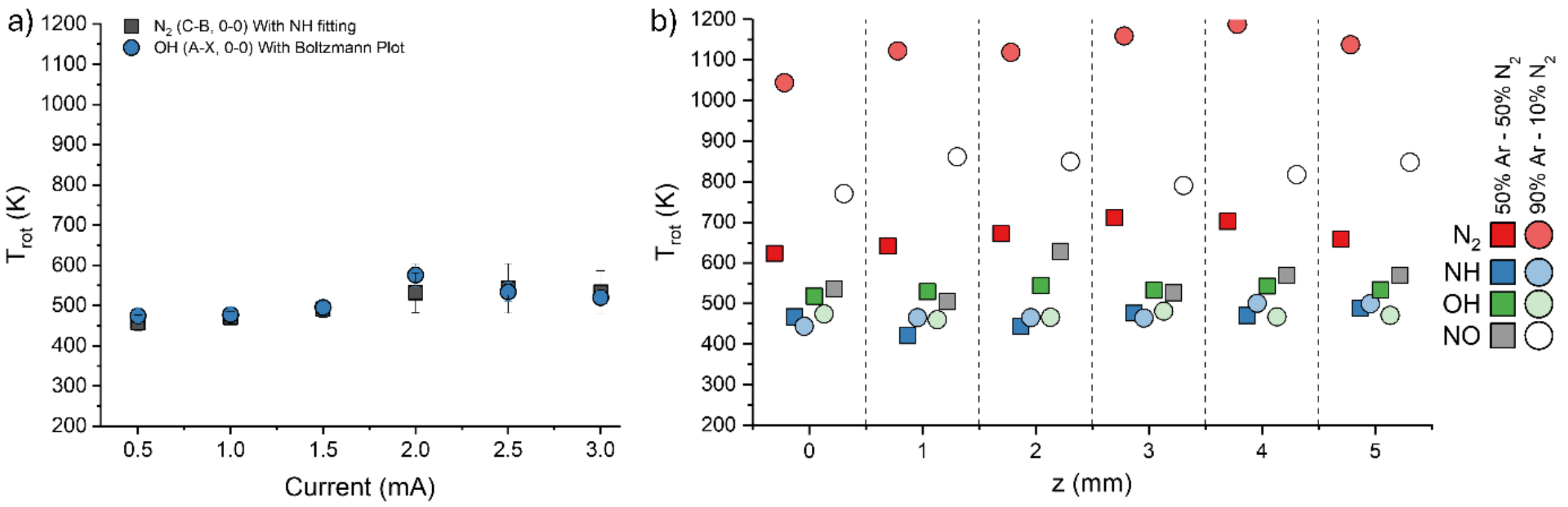

**Fig.13**. **a**) Comparison of determined $T_{rot}$ of OH(A) (obtained using the single $T_{rot}$ fitting and Boltzmann plot methods) with the $T_{rot}$ of $N_2$(C) obtained through fitting of the $N_2$(SPS) transitions around 337 (after deconvolution of NH(A-X) emission), 357 and 380 nm. The conditions are for $P_{high}$=30 mbar, $P_{low}$=10 mbar, 50%-50% Ar-$N_2$. These measurements were conducted in triplicate, with the resulting error bars representing these repetitions. **b**) Comparison of determined $T_{rot}$ of OH(A) (obtained using the Boltzmann plot method), NO(A), NH(A) and $N_2$(C) (after deconvolution of NH(A-X) emission), obtained through fitting of the transitions around 237, 336 and 337 nm, respectively, along the jet axis. The conditions are for $I_d$=2.5 mA, $P_{high}$=30 mbar, $P_{low}$=10 mbar and gas mixtures of 50%-50% Ar-$N_2$ and 90%-10% Ar-$N_2$.

In the literature, several approaches have been used to obtain $T_{Gas}$ in MHCDs. OES allows analyzing the rotational distribution of molecular bands and fitting atomic line profiles broadened by collisions. For instance, Lazzaroni et al. [46] showed that in DC argon MHCDs at 30–200 Torr, the $T_{rot}$ of the $N_2$(B) state can offer a good estimate of the $T_{Gas}$ being about 450–500 K. In MHCDs containing Helium at 100 Torr, Dufour et al. [47] used trace amounts of $N_2$ and demonstrated that the $T_{rot}$ of $N_2$(C) reliably approximates the $T_{Gas}$ for different discharge currents. At atmospheric pressure conditions, Sismanoglu et al. [4] compared $T_{rot}$ of OH(A) inferred from the OH(A–X) band with Ar I line (603 nm) broadening and found a good agreement between the two methods (460–640 K in the current range 7–15 mA). This also validates the use of OH(A) in our case as a valid thermometric probe and shows that cross-validation of different methods is important to enhance accuracy. Other works have adopted atomic line-shape thermometry (via broadening analysis of He I and Ar I resonant lines) [22, 23] to infer $T_{Gas}$ in near atmospheric pressure MHCD. Finally, in ns-pulsed Ar/$N_2$ MHCD arrays, Kasri et al. [48] reported time-averaged $N_2$(C) rotational temperatures suggesting that afterglow $N_2$(C) excitation can affect the $T_{rot}$, making it an overestimation of $T_{Gas}$. A critical limitation in Ar/$N_2$ mixtures is the influence of $Ar^m$. As demonstrated by Iseni et al. [22], in argon-rich MHCD, argon metastables can non-thermally populate $N_2$(C), yielding different $T_{rot}$ values for different vibrational bands and leading to systematic overestimates of $T_{Gas}$. This effect is further complicated in pulsed regimes, where Kasri et al. (2019) [48] reported that afterglow excitation processes also lead to overestimated time-averaged $T_{rot}$ values. Consequently, relying on a single emission band without a precaution can yield misleading results when the MHCD operates across a wide range of gas mixtures and pressures.

Despite the wealth of published studies, they often employ isolated approaches to infer $T_{Gas}$, i.e., analysis of emission profiles from one probe molecule or atom generated in MHCD. Although this choice strongly depends on the MHCD conditions and the availability of probe species, a cross-comparison (when possible) of different approaches to infer $T_{Gas}$ is highly desirable. The present work addresses these limitations by studying/examining the spatial variation of the $T_{rot}$ of multiple molecular bands (NO(A), OH(A), NH(A), $N_2$(C)) in an Ar/$N_2$ MHCD jet operating in various gas mixtures and pressures, with the goal of identifying the most suitable molecular probe for accurate determination of $T_{Gas}$ under the considered experimental conditions.

## CONCLUSIONS

In this work, the emission properties of a DC-driven Ar/$N_2$ MHCD jet were investigated under various operating conditions: pressure, discharge current, gas mixture. The morphology of the jet and the cathodic expansion were studied for different currents and Ar/$N_2$ mixtures using a conventional camera. An asymmetric cathodic expansion was revealed in $N_2$-rich mixtures compared to pure Ar operation. The plasma is essentially following the most conductive point on the cathode surface. The MHCD jet is longer and brighter in $N_2$-rich mixtures possible due to enhanced reactivity induced by Ar and/or nitrogen metastables in the discharge afterglow.

Furthermore, a systematic assessment of the emissions from several molecular bands, $N_2$(C–B), OH(A–X), NO(A–X) and NH(A–X), as potential $T_{Gas}$ probes was performed in a DC Ar/$N_2$ MHCD jet. Rotational and vibrational temperatures were obtained by either fitting measured spectra with synthetic spectra or by using the Boltzmann plot method. Across the range of all conditions studied, the discharge is clearly non-equilibrium: $T_{vib}$ of $N_2$(C) remain in the 4.000–5.000 K range, while $T_{rot}$ are in the 350–650 K range.

For a 50%Ar-50%$N_2$ mixture, the $T_{rot}$ inferred from $N_2$(C) and OH(A) (and, where detectable, NH(A)) are in relatively good mutual agreement and show only modest axial variation. In this regime, all three probes yield $T_{rot}$ values that increase with discharge current, as expected, and remain relatively uniform along the jet, supporting their use as reliable surrogates for $T_{Gas}$. These observations indicate that, under conditions where metastable-driven excitations are not dominant, commonly used $N_2$(C-B) emission around 337 nm can still be used as a robust temperature probe, particularly when its spectra are carefully deconvolved from overlapping bands such as NH(A-X).

In contrast, for Ar-rich mixtures, $N_2$(C) and NO(A) no longer provide straightforward access to $T_{Gas}$. Their apparent rotational temperatures can exceed 800–1100 K, significantly above the values deduced from OH(A) (≈450–650 K), even though the discharge operating conditions are similar. This behaviour is consistent with enhanced non-thermal population of the upper electronic states via collisions with $Ar^m$ and $N_2$(A), which affects the rotational distributions and leads to a systematic overestimation of the actual $T_{Gas}$. OH(A), and NH(A) (when detectable), appear much less sensitive to these effects.

Overall, it was shown that OH(A) is the most trustworthy molecular probe for gas temperature in this MHCD configuration, with NH(A) also providing a valuable cross-check when present. $N_2$(C) and NO(A) can only be used with caution, primarily in $N_2$-rich or intermediate mixtures where metastable-driven excitations are limited. The present work highlights the importance of cross-validating multiple molecular probes and considering excitation kinetics when interpreting $T_{rot}$ in metastable-rich microplasmas and provide a valuable guide for the assessment of $T_{Gas}$ in Ar/$N_2$ environments.

## Acknowledgments and Funding Information

This work was supported by the French Research National Agency (ANR) via the project SPECTRON (No. ANR-23-CE51-0004-01), PlasBoNG (No. ANR-20-CE09-0003-01), DESYNIB (No. ANR-16-CE08-0004), and LabEx SEAM (Nos. ANR-10-LABX-0096 and ANR-18-IDEX-0001). The authors would like to thank Noel Girodon-Boulandet for technical drawing and support.

## Author contributions

**Dimitrios Stefas**: Conceptualization (supporting); Methodology (supporting); Software (lead); Validation (lead); Formal analysis (lead); Investigation (lead); Data Curation (lead); Writing - Original Draft (lead); Writing - Review & Editing (equal); Visualization (lead).
**Belkacem Menacer**: Investigation (supporting); Data Curation (supporting); Writing - Original Draft (supporting); Writing - Review & Editing (equal); Visualization (supporting).
**Nikolaos Chazapis**: Investigation (supporting); Data Curation (supporting); Writing - Review & Editing (equal); Visualization (supporting).
**Alice Remigy**: Conceptualization (supporting); Investigation (supporting); Data Curation (supporting); Writing - Review & Editing (equal); Visualization (supporting).
**Guillaume Lombardi**: Conceptualization (supporting); Methodology (supporting); Validation (supporting); Resources (equal); Writing - Review & Editing (equal); Visualization (supporting); Supervision (equal); Project administration (supporting); Funding acquisition (equal).
**Claudia Lazzaroni**: Conceptualization (supporting); Methodology (supporting); Validation (supporting); Resources (equal); Writing - Review & Editing (equal); Visualization (supporting); Supervision (equal); Project administration (supporting); Funding acquisition (equal).
**Kristaq Gazeli**: Conceptualization (lead); Methodology (lead); Software (supporting); Validation (lead); Investigation (supporting); Resources (equal); Data Curation (supporting); Writing - Original Draft (supporting); Writing - Review & Editing (equal); Visualization (supporting); Supervision (equal); Project administration (lead); Funding acquisition (equal).

## Statements and Declarations

### Competing interests

The authors declare no competing interests.

### Data Availability

Data are available from the corresponding author upon reasonable request